\documentclass[fleqn,usenatbib]{mnras}

\usepackage{newtxtext,newtxmath}
\usepackage[T1]{fontenc}

\DeclareRobustCommand{\VAN}[3]{#2}
\let\VANthebibliography\thebibliography
\def\thebibliography{\DeclareRobustCommand{\VAN}[3]{##3}\VANthebibliography}

\usepackage{graphicx}	
\usepackage{amsmath}	

\usepackage{makecell}
\usepackage{multirow}
\usepackage{ctable}
\usepackage{ulem}  

\newcolumntype{?}{!{\vrule width 2pt}}

\newlength{\Oldarrayrulewidth}
\newcommand{\Cline}[2]{%
  \noalign{\global\setlength{\Oldarrayrulewidth}{\arrayrulewidth}}%
  \noalign{\global\setlength{\arrayrulewidth}{#1}}\cline{#2}%
  \noalign{\global\setlength{\arrayrulewidth}{\Oldarrayrulewidth}}}
\definecolor{darkpastelgreen}{rgb}{0.01, 0.75, 0.24}

\title[DeepMerge II: Building Robust Deep Learning Algorithms for Merging Galaxy Identification Across Domains]{DeepMerge II: Building Robust Deep Learning Algorithms for Merging Galaxy Identification Across Domains}

\author[A. \'Ciprijanovi\'c et al.]{
A. \'Ciprijanovi\'c,$^{1}$\thanks{E-mail: aleksand@fnal.gov}
D. Kafkes,$^{1}$
K. Downey,$^{2}$
S. Jenkins,$^{2}$
G. N. Perdue,$^{1}$
S. Madireddy,$^{3}$
T. Johnston,$^{4}$\newauthor
G. F. Snyder,$^{5}$
B. Nord$^{1,2,6}$
\\
$^{1}$Fermi National Accelerator Laboratory, P.O. Box 500, Batavia, IL 60510, USA\\
$^{2}$Department of Astronomy and Astrophysics, University of Chicago, IL 60637, USA\\
$^{3}$Argonne National Laboratory, 9700 S Cass Ave, Lemont, IL 60439, USA\\
$^{4}$Oak Ridge National Laboratory, 1 Bethel Valley Rd, Oak Ridge, TN 37830, USA\\
$^{5}$Space Telescope Science Institute, 3700 San Martin Drive, Baltimore, MD 21218, USA\\
$^{6}$Kavli Institute for Cosmological Physics, University of Chicago, Chicago, IL 60637, USA
}

\definecolor{darkbluegrey}{rgb}{0.45, 0.43, 0.66}
\definecolor{vermillion}{rgb}{0.86, 0.18, 0.01}

\date{Accepted XXX. Received YYY; in original form ZZZ}

\pubyear{2021}

\begin{document}
\label{firstpage}
\pagerange{\pageref{firstpage}--\pageref{lastpage}}
\maketitle

\begin{abstract}
In astronomy, neural networks are often trained on simulation data with the prospect of being used on telescope observations. Unfortunately, training a model on simulation data and then applying it to instrument data leads to a substantial and potentially even detrimental decrease in model accuracy on the new target dataset. Simulated and instrument data represent different data domains, and for an algorithm to work in both, domain-invariant learning is necessary. Here we employ domain adaptation techniques--- Maximum Mean Discrepancy (MMD) as an additional transfer loss and Domain Adversarial Neural Networks (DANNs)--- and demonstrate their viability to extract domain-invariant features within the astronomical context of classifying merging and non-merging galaxies. Additionally, we explore the use of Fisher loss and entropy minimization to enforce better in-domain class discriminability. We show that the addition of each domain adaptation technique improves the performance of a classifier when compared to conventional deep learning algorithms. We demonstrate this on two examples: between two Illustris-1 simulated datasets of distant merging galaxies, and between Illustris-1 simulated data of nearby merging galaxies and observed data from the Sloan Digital Sky Survey. The use of domain adaptation techniques in our experiments leads to an increase of target domain classification accuracy of up to ${\sim}20\%$. With further development, these techniques will allow astronomers to successfully implement neural network models trained on simulation data to efficiently detect and study astrophysical objects in current and future large-scale astronomical surveys.

\end{abstract}




\section{Introduction}
\label{sec:intro}

Studies of galaxy mergers are crucial for understanding the evolution of galaxies as astronomical objects, their star formation rates, chemistry, particle acceleration and other properties. Moreover, they are equally important for cosmology, understanding structure formation and the study of evolution of matter in the universe. Being able to leverage large samples of merging galaxies and to connect the knowledge obtained from large-scale simulations and astronomical surveys will play an important role in these studies. \\
\indent Standard methods for classifying merging galaxies using visual inspection~\citep{LK2004} or extraction of parametric measurements of structure such as the S\'ersic index~\citep{S1963}, Gini coefficient, M20 -- the second-order moment of the brightest $20\%$ percent of the galaxy’s flux~\citep{LP2004}, CAS -- Concentration, Asymmetry, Clumpiness~\citep{CB2003} etc., can be time consuming, prone to biases, or require the use of high-quality images. Due to these limitations, it has been shown that machine learning can greatly advance the study of merging galaxies ~\citep{AS2018,SR2019,PW2019,CS2020}, improving both the quality of the results and the speed of working with big datasets. Studies of merging galaxies using machine learning can greatly benefit from models trained on simulated data, which can then be successfully applied to newly observed images from present and future large-scale surveys. \\ 
\indent Simulated and observed images have different origins and represent different data domains. In this case, labeled simulation images represent the source domain we are starting from, while observed data (often unlabeled) is the target domain. While images produced by simulations are made to mimic real observations from a particular telescope, unavoidable small differences can cause the model trained on simulated images to perform substantially worse when applied to real data. This substandard performance has been demonstrated directly in the case of merging galaxies by~\citet{CS2020}. The authors show that even when the only difference between the two merging galaxy datasets is the inclusion of noise, convolutional neural networks (CNNs) trained on one dataset cannot classify the other dataset at all. In the paper the classification accuracy of in-domain images was $79\%$, while the accuracy for out-of-domain images was around $50\%$, equivalent to random guessing. Additionally,~\cite{PW2019} use a dataset from the EAGLE simulation~\citep{SC2015}, which is made to mimic Sloan Digital Sky Survey (SDSS) observations, and real SDSS images~\citep{LS2008,DK2010}. Their work provides further evidence that the performance of the classifier trained on one dataset has much lower accuracy when classifying the other dataset. They achieve out-of-domain accuracies between $53-65\%$, with the classifier trained on real SDSS images classifying EAGLE simulation images performing particularly poorly. These two examples are indicative of a great need for more sophisticated deep learning methods to be applied to cross-domain studies in astrophysical contexts. \\
\indent An important area of deep learning research includes the development of Domain Adaptation (DA) techniques~\citep{C2017,WD2018,GD2020}. They allow the model to learn the invariant features shared between the domains and align the extracted latent feature distributions. This allows the model to successfully find the decision boundary that distinguishes between different classes in multiple domains at the same time. One group of divergence-based DA methods includes finding and minimizing some divergence criterion between the source and target data distributions. Some of the most well known methods include Maximum Mean Discrepancy  (MMD;~\cite{GB2008}), Correlation Alignment (CORAL;~\cite{SS2016,SF2016}), Contrastive Domain Discrepancy (CDD;~\cite{KJ2019}), and the Wasserstein metric~\citep{SQ2018}. On the other hand, adversarial-based DA methods use either generative models~\citep{LT2016} to create synthetic target data related to the source domain or more simple models that utilize domain-confusion loss~\citep{GU2016}, which measures how well the model distinguishes between different data domains.\\
\indent In this paper we employ two different domain adaptation techniques--- Maximum Mean Discrepancy (MMD) and domain adversarial training with a domain-confusion loss--- to improve cross-domain applications of deep learning models to the problem of distinguishing between merging and non-merging galaxies. Maximum Mean Discrepancy works by minimizing a distance measure of the mean embeddings of the two domain distributions in latent feature space~\citep{GB2008}. It is applied to standard classification networks as a transfer loss. Domain adversarial neural networks (DANNs; ~\cite{GU2016}), use adversarial training between a label classifier--- which distinguishes mergers from non-mergers--- and a domain classifier--- which classifies the source and target domain of images. This kind of training employs a gradient reversal layer within the domain classifier, thereby maximizing the loss in this branch and leading to the extraction of domain-invariant features from both sets of images. Following methods from~\citet{ZZ2020}, we also add Fisher loss and entropy minimization~\citep{GB2004}, which can be used as additional losses for both MMD or domain adversarial training, to improve the overall performance of the classifier. Both of these loss functions enforce additional discriminability between the classes in source (Fisher loss) and target (entropy minimization) domains, by producing more compact classes in the latent feature space. \\
\indent We test two networks for a comparison of technique results across architectures: DeepMerge, a simple convolutional network for classification of galaxies presented in~\citet{CS2020}, as well as the more complex and well-known ResNet18~\citep{HZ2015}. We demonstrate both methods on a dataset similar to the one from~\citet{CS2020}, using simulated distant merging galaxies from Illustris-1~\citep{VG2014} at redshift z = 2, both without (source) and with (target) the addition of random sky shot noise to mimic observations from the Hubble Space Telescope. \\
\indent Additionally, we test these methods on a harder and more realistic application example, where the source domain includes simulated galaxies at $z=0$ from the Illustris-1 simulation~\citep{VG2014} made to mimic SDSS observations and real SDSS images of merging galaxies~\citep{LS2008,DK2010}. These two domains exhibit a much larger discrepancy, and simply applying MMD and adversarial training does not perform well. We demonstrate that combining MMD with transfer learning from the model trained on the first dataset of distant merging galaxies can be used to solve this harder domain adaptation problem.\\
\indent With the use of domain adaptation techniques mentioned above, we manage to increase the target domain classification accuracy up to ${\sim}20\%$ in our experiments, which allowed the model to be successfully used in both domains. It is our hope that the use and continued development of these techniques will allow astronomers and cosmologists to develop deep learning algorithms that can combine information from either simulations and real data, or to combine observations from different telescopes. \\
\indent The remainder of the paper is structured as follows: In Section~\ref{sec:methods}, we introduce and explain the domain adaptation methods used in this paper. We explain the neural network architectures we use in Section~\ref{sec:networks}. In Section~\ref{sec:data}, we give details about the images we use in our experiments and talk more about the experimental setup in Section~\ref{sec:experiments}. Finally, our results are given in Section~\ref{sec:results}, followed by a discussion in Section~\ref{sec:discuss}.

\section{Methods}
\label{sec:methods}
Deep learning is already bringing advances in astronomy and survey science, as in other academic fields and industry. Many astronomical applications often require these models to perform well on new datasets, requiring the applicability of features learned from simulations to data that the model was not initially trained on, including newly available observed data and cross-telescope applications. Since labelling new data is slow and prone to errors, retraining these neural networks on new datasets in order to maintain high performance is often impractical. In these situations, a discriminative model that is able to transfer knowledge between training (source domain) and new data (target domain) is necessary. This can be achieved by using domain adaptation techniques, which extract invariant features between two domains, so that a neural network classifier trained on the source domain can also be applied successfully on a target domain. As previously underscored, this functionality is very useful in situations often found in astronomy, where the target domain is comprised of new observational data that has very few identified objects or is completely unlabeled. Here we will test several DA techniques that can be effective in the situation where the target domain is unlabeled. These techniques include adding transfer loss to the widely-adopted cross-entropy loss used for standard classification of images.

Cross-entropy loss is given as:
\begin{equation}
{\cal L_\mathrm{CL}} = - \sum\limits_{m=1}^{\mathrm{M}} y_m \log p(y_m),
\end{equation}
where $m$ is a particular class and the total number of classes is $\mathrm{M}$. The true label for class $m$ is given as $y_m$, and $p(y_m)$ is the neural network assigned score, i.e. the output prediction from the last layer, for a given class $m$.
Minimizing cross-entropy loss leads to output predictions approaching real label values, which results in an increase in classification accuracy.

The inclusion of transfer loss allows DA techniques to impact the way the network learns via backpropagation. We explore two different transfer losses in this paper: Maximum Mean Discrepancy (MMD; ~\cite{GS2012}) and using the loss of the discriminator from a Domain Adversarial Neural Network (DANN; ~\cite{GU2016}). Additionally, we explore adding Fisher loss~\citep{ZZ2020}, which enforces feature discriminability between classes in the source domain, and entropy minimization, which forces a target sample to move toward one of the compacted and separated source classes~\citep{GB2004}, leading to better class separability in the target domain.

The resultant total classifier loss~\citep{ZZ2020} has multiple components:
\begin{equation}
{\cal L}_\mathrm{TOT} = {\cal L}_\mathrm{CL} + \lambda_\mathrm{FL}{\cal L}_\mathrm{FL} + \lambda_\mathrm{EM}{\cal L}_\mathrm{EM} + \lambda_\mathrm{TL}{\cal L}_\mathrm{TL},\label{eq:total}
\end{equation}
where we define ${\cal L}_\mathrm{TOT}$, ${\cal L}_\mathrm{CL}$, ${\cal L}_\mathrm{FL}$, ${\cal L}_\mathrm{EM}$, ${\cal L}_\mathrm{TL}$ as total loss, classifier loss, Fisher loss, entropy minimization, and transfer loss, respectively. The contribution of these additional losses can be weighted using weights $\lambda_\mathrm{FL}$, $\lambda_\mathrm{EM}$ and $\lambda_\mathrm{TL}$. Further details about the different losses used are given below.

\subsection{Transfer Loss}

The transfer loss ${\cal L}_\mathrm{TL}$ is calculated from the DA technique, whose goal is to decrease discrimination between the source and the target domains. This involves the representation of data from both domains in a higher-dimensional latent feature space. In this paper, we explore the use of both MMD and domain adversarial training as transfer criteria. MMD frames the domain problem in terms of high-dimensional statistics and involves calculating the distance between the mean embeddings of the source and target domain distributions. In the case of adversarial training, the domain discrepancy problem is addressed by adopting DANN, a neural network that seeks to find the common feature space between the source and target distributions by jointly minimizing the training loss in the source domain while maximizing the loss of the domain classifier.

We will denote the source and target domains as ${\cal D}_\mathrm{s}$ and ${\cal D}_\mathrm{t}$ respectively. Source domain images are labeled, so we have $n_\mathrm{s}$ pairs of images and labels $\{\mathbf{x}_\mathrm{s}, \mathbf{y}_\mathrm{s}\}$, while in the case of the target domain we have $n_\mathrm{t}$ unlabeled images $\mathbf{x}_\mathrm{t}$. Images from both domains are associated with domain labels $\mathbf{d}_\mathrm{s}$ for source domain and $\mathbf{d}_\mathrm{t}$ for target domain.

\subsubsection{Maximum Mean Discrepancy (MMD)}

Maximum Mean Discrepancy (MMD) is a statistical technique that calculates a nonparametric distance between mean embeddings of the source and target probability distributions from the $L_{\infty}$ norm. Following ~\citet{SG2007} and ~\citet{GB2008}, we designate the source probability distribution as ${\cal P}_\mathrm{s}$ and the target probability distribution as ${\cal P}_\mathrm{t}$.

It is possible to estimate densities of ${\cal P}_\mathrm{s}$ and ${\cal P}_\mathrm{t}$ from the observed source and target data using kernel methods, but this estimation, which is often computationally expensive and introduces bias, is unnecessary in practice~\citep{JI2010}. Instead, we use kernel methods to determine their means for subtraction instead of estimating the full distributions:

\begin{equation}
\label{eq:mmd}
    D( {\cal P}_\mathrm{s}, {\cal P}_\mathrm{t}, {\cal F}) := \sup_{f \in {\cal F}}\mathbb{E}_{{\cal P}_\mathrm{s}}[f(x_\mathrm{s})] - \sup_{f \in {\cal F}}\mathbb{E}_{{\cal P}_\mathrm{t}}[f(x_\mathrm{t})],
\end{equation}

\noindent where $D$ denotes the kernel distance as a proxy for discrepancy, $x_\mathrm{s}$ and $x_\mathrm{t}$ are random variables drawn from ${\cal P}_\mathrm{s}$ and ${\cal P}_\mathrm{t}$ respectively, function class ${\cal F}$ closely resembles the set of CDF functions in vector space with total variance less than one operating on the domain $(-\infty$, ${\rm I\!R}]$ and the supremum is the least element in ${\cal F}$ greater than or equal to the chosen $f$, i.e. the max of the subset. By this definition, if ${\cal P}_\mathrm{s}$ = ${\cal P}_\mathrm{t}$, then $D( {\cal P}_\mathrm{s}, {\cal P}_\mathrm{t}, {\cal F}) = 0$~\citep{GS2012, JI2010}. If ${\cal P}_\mathrm{s} \neq {\cal P}_\mathrm{t}$, then there must exist some $f$ such that the distance between the two means is maximized. This becomes an optimization problem in which the criterion aims to maximize the discrepancy by separating the two distributions as far as possible in some high-dimensional feature space. Kernel methods are well suited to this task since they are able to map means into higher-dimensional Reproducing Kernel Hilbert Spaces (RKHSs). Furthermore, this embedding linearizes the metric by mapping the input space into a feature vector space.

RKHSs possess several properties that facilitate the calculation of $D$, including a property of norms that rescale the output of a function to fit within a unit ball--- which greatly restricts the many possibilities of function class ${\cal F}$--- and a reproducing property that reduces calculations to the inner product of the output of functions. For example,  $\langle f, k(x, \cdot) \rangle = f(x)$, where $k(x, \cdot)$ is a kernel that has one argument fixed at $x$, and the second free. Performing this calculation in an RKHS means there is no need to explicitly calculate the mapping function $\phi(x)$ that maps $x_\mathrm{s}$ and $x_\mathrm{t}$ into an RKHS feature space due to equivalence between $\phi(x)$ and $k(x,\cdot) : \langle \phi(x), \phi(x') \rangle = \langle k(x,x') \rangle = \langle k(x, \cdot), k(x', \cdot) \rangle$. Therefore, Eq.~\ref{eq:mmd} can be re-expressed as two inner products with kernels mean embeddings in an RKHS:
\begin{equation}
\label{eq:mmdrkhs}
    D( {\cal P}_\mathrm{s}, {\cal P}_\mathrm{t}, {\cal F}) := \sup_{||f|| \leq 1}\mathbb{E}_{{\cal P}_\mathrm{s}}[\langle k(x_\mathrm{s}), f \rangle] - \sup_{||f|| \leq 1}\mathbb{E}_{{\cal P}_\mathrm{t}}[\langle k(x_\mathrm{t}), f \rangle]
\end{equation}

\hspace{15.mm} $= \sup_{f}\langle \mu_\mathrm{s}, f \rangle - \sup_{f}\langle \mu_\mathrm{t}, f \rangle = \sup_{f}\langle \mu_\mathrm{s} - \mu_\mathrm{t}, f \rangle$,\\

\noindent where $\mu_\mathrm{s}$ and $\mu_\mathrm{t}$ are the source and target distribution's mean embeddings and $f$ is still bounded by the unit ball of the RKHS. Clearly, the inner product is maximized for the identity $\langle a,a \rangle = 1$. Therefore, to maximize the mean discrepancy we need $f=\mu_\mathrm{s}$ - $\mu_\mathrm{t}$, leaving us with the final formula:
\begin{equation}
   D( {\cal P}_\mathrm{s}, {\cal P}_\mathrm{t}, {\cal F}) :=  \mathbb{E}_{{\cal P}_\mathrm{s}}[k(x_\mathrm{s},x'_\mathrm{s})] - 2 \mathbb{E}_{{\cal P}_\mathrm{s,t}}[k(x_\mathrm{s},x_\mathrm{t})] + \mathbb{E}_{{\cal P}_\mathrm{t}}[k(x_\mathrm{t},x'_\mathrm{t})],
\end{equation}

\noindent where all kernel functions come from the simplification of the inner product $\langle k(x, \cdot), k(x', \cdot) \rangle$, following the logic of the equivalence between the mapping function and the kernel established previously. Here it is clear that the distance is expressed as the difference between the self-similarities of source and target domains and their cross-similarity.

In practice, this is discretized to give the the unbiased estimator $D$:

\begin{equation}
D = \frac{1}{q(q-1)} \sum\limits_{i!=j} k(x_\mathrm{s}(i), x_\mathrm{s}(j)) - k(x_\mathrm{s}(i), x_\mathrm{t}(j))-
\end{equation}
\hspace{25mm} $ k(x_\mathrm{t}(i), x_\mathrm{s}(j)) + k(x_\mathrm{t}(i), x_\mathrm{t}(j))$,\\

\noindent where $q$ is the sample number of $x_\mathrm{s}$ and $x_\mathrm{t}$. While in practice, $k$ can be considered a general kernel, we follow ~\citet{ZZ2020} and substitute $k$ with $\kappa$, where $\kappa$ is a linear combination of multiple Gaussian Radial Basis Function (RBF) kernels to extend across a range of mean embeddings. Gaussian RBF kernel can be written as:

\begin{equation}
k(x,x')=e^{-\frac{||x-x'||^2}{2\sigma^2}},
\end{equation}
where $||x-x'||$ is the Euclidean distance norm (where $x$ can be $x_\mathrm{s}$ or $x_\mathrm{t}$ depending on the domain), and $\sigma$ is the free parameter which determines the width of the kernel.

Finally, we use MMD as our transfer loss: ${\cal L}_\mathrm{TL,mmd} = D$, effectively drawing the source and target distributions together in latent space as the network aims to minimize the loss via backpropagation.

\subsubsection{Domain Adversarial Training}

Domain adversarial training employs a Domain Adversarial Neural Network (DANN) to distinguish between the source and target domains~\citep{GU2016}. DANNs are comprised of three parts: a feature extractor ($N_\mathrm{F}$), label predictor ($N_\mathrm{L}$), and domain classifier ($N_\mathrm{D}$). The first two parts can be found in any Convolutional Neural Network (CNN)--- the feature extractor is built from convolutional layers which extract features from images, while the label predictor usually has fully-connected (dense) layers which output the class label. The last part, the domain classifier, is unique to DANNs. It is built from dense layers and optimized to predict the domain labels. This domain classifier is added after the feature extractor as a parallel branch to the label predictor. It includes a gradient reversal layer which maximizes the loss for this branch of the neural network, thus achieving the adversarial objective of confusing the discriminator. When the domain classifier fails to distinguish latent features from the two domains, the domain invariant features, i.e. the shared feature space, is found.

Compared to regular CNNs which can learn the best features for classification, training DANNs can lead to a slight drop in classification accuracy for the source domain because only the domain-invariant features are used. However, this will also lead to an increase in the classification accuracy in the new target domain which is our objective. The total loss for a DANN is ${\cal L}_\mathrm{DANN} = {\cal L}_\mathrm{class} - \lambda{\cal L}_\mathrm{d}$, where ${\cal L}_\mathrm{class}$ is the loss for the image class label predictor $N_\mathrm{L}$, while ${\cal L}_\mathrm{d}$ is the loss from the domain classifier $N_\mathrm{D}$. Fine-tuning the trade-off between these two quantities during the learning process is done with the regularization parameter $\lambda$. Domain classifier loss is calculated as:
\begin{equation}
{\cal L}_\mathrm{d} = \frac{1}{n_\mathrm{s}} \sum\limits_{\mathbf{x}_\mathrm{s}\in{\cal D}_\mathrm{s}} l(N_\mathrm{D}(N_\mathrm{F}(\mathbf{x}_\mathrm{s})),d_\mathrm{s})   + \frac{1}{n_\mathrm{t}}  \sum\limits_{\mathbf{x}_\mathrm{t}\in{\cal D}_\mathrm{t}} l(N_\mathrm{D}(N_\mathrm{F}(\mathbf{x}_\mathrm{t})),d_\mathrm{t}),
\end{equation}
where $l(N_\mathrm{D}(N_\mathrm{F}(\mathbf{x}_\mathrm{s})),d_\mathrm{s})$ and $l(N_\mathrm{D}(N_\mathrm{F}(\mathbf{x}_\mathrm{t})),d_\mathrm{t})$ are the output scores for the source domain and target domain labels, respectively. Similarly to the class label predictor, the output scores for domain labels are also calculated using cross-entropy loss on domain labels.
Finally, we can use domain classifier loss as our transfer loss:
\begin{equation}
{\cal L}_\mathrm{TL,adv} = \mathrm{max} \{ -{\cal L}_\mathrm{d} \}.
\end{equation}

\subsection{Fisher Loss}
The addition of Fisher loss to the classification and transfer losses was demonstrated to further improve classification performance for source domain images in ~\citet{ZZ2020}. This improvement in source classification can aid the performance of both MMD and domain adversarial training transfer criteria. It is more generally applicable than Scatter Component Analysis~\citep{GB2015}, which also results in class compactness, but can only be used in conjunction with MMD and would not be practical for use with adversarial training methods ~\citep{ZZ2020}. 

Minimizing Fisher loss leads to within-class compactness and between-class separability in the latent feature space, which makes the distinction between classes easier in the source domain. Fisher loss produces a centroid for each class and effectively pushes labeled classes toward their respective centroids, thereby creating more tightly clustered classes further apart from each other. It can be defined as a function $f$:
\begin{equation}
{\cal L}_\mathrm{FL} = f(\mathrm{tr}(\mathbf{S}_\mathrm{w}),\mathrm{tr}(\mathbf{S}_\mathrm{b}))).
\label{eq:fisher}
\end{equation}

Here, $\mathbf{S}_\mathrm{w} = \sum_{m=1}^\mathrm{M} \sum_{j=1}^{n_\mathrm{m}} (\mathbf{h}_{m_j} - \mathbf{c}_m) (\mathbf{h}_{m_j} - \mathbf{c}_m)^\mathrm{T}$ captures the intra-class dispersion of samples within each class, where $\mathbf{h}_{m_j}$ is the latent feature of the $j$-th sample of the $m$-th class (with M being the total number of classes). On the other hand,  $\mathbf{S}_\mathrm{b} = \sum_{m=1}^\mathrm{M} (\mathbf{c}_m-\mathbf{c}) (\mathbf{c}_m-\mathbf{c})^\mathrm{T}$ describes the distances of all class centroids $\mathbf{c}_m$ to the global center $\mathbf{c}=\frac{1}{\mathrm{M}}\sum_{m=1}^{\mathrm{M}} \mathbf{c}_m$. This global center, $\mathbf{c}$, is meant to be optimized such that the centroids of the classes are pushed as far away as possible. Traces are used in the computation of the Fisher loss since they are computationally efficient.

To achieve the intended result of intra-class compactness and inter-class separability, Eq.~\ref{eq:fisher} must be monotonically increasing with respect to the trace of the intra-class matrix $\mathbf{S}_\mathrm{w}$ and monotonically decreasing with respect to the trace of the inter-class matrix $\mathbf{S}_\mathrm{b}$. Thus, as the loss is minimized via backpropagation, the distances within classes will grow smaller and the distances between classes will grow larger. There are two simple ways one can construct a Fisher loss function obeying these constraints: Fisher trace ratio ${\cal L}_\mathrm{FL} = \mathrm{tr}(\mathbf{S}_\mathrm{w}) / \mathrm{tr}(\mathbf{S}_\mathrm{b})$) or Fisher trace difference ${\cal L}_\mathrm{FL} = \mathrm{tr}(\mathbf{S}_\mathrm{w}) - \mathrm{tr}(\mathbf{S}_\mathrm{b})$). In this paper, we have chosen to use the Fisher trace ratio as our Fisher loss.

As we mentioned in Eq.~\ref{eq:total} for total loss, we can control the contribution of all additional losses using the weight parameter $\lambda$. In the case of Fisher loss, we can separately weight the importance of the two matrices using $\lambda_\mathrm{w}$ and $\lambda_\mathrm{b}$. Since we have chosen to use the trace ratio Fisher loss ${\cal L}_\mathrm{FL} = \lambda_\mathrm{w}\mathrm{tr}(\mathbf{S}_\mathrm{w}) / \lambda_\mathrm{b}\mathrm{tr}(\mathbf{S}_\mathrm{b})$, this gives $\lambda_\mathrm{FL}=\lambda_\mathrm{w}/\lambda_\mathrm{b}$.  

\subsection{Entropy Minimization}
Fisher loss can only be used in the source domain since it requires ground-truth labels to calculate intra-class centroids and the between class global center. However, Fisher loss can aid the discrimination between classes within the unlabeled target domain as well through entropy minimization loss. Entropy minimization loss pushes examples from the target domain toward source domain class centroids. Therefore, entropy minimization ensures better generalization of the decision boundary between optimally discriminative and compact source domain classes to the target domain as well~\citep{GB2004}.

Entropy minimization loss is defined as:
\begin{equation}
{\cal L}_\mathrm{EM} = - \sum\limits_{j=1}^{n_\mathrm{m}} \sum\limits_{m=1}^\mathrm{M} p(y_m|\mathbf{h}_{m_j}) \log p(y_m|\mathbf{h}_{m_j}),
\end{equation}

\noindent where $p(y_m|\mathbf{h}_{m_j})$ is the classifier output and the true label is not needed. The above formula is based on Shannon's entropy~\citep{S1948}, which for a discrete probability distribution can be written as $H(x)=-\sum_{i=1}^{\mathrm{N}} p_i \log_2 p_i$.

\section{Neural network architectures}
\label{sec:networks}

We present the performance of domain adaptation using the aforementioned techniques in two neural networks--- the simpler DeepMerge architecture with 174,626 trainable parameters~\citep{CS2020} and the more complex and well known network ResNet18 with 22,484,866 trainable parameters~\citep{HZ2015}--- to compare results across architectures. We decided to use the smallest standard ResNet  architecture, in order to more easily tackle possible overfitting of the model, due to small sizes of merging galaxies datasets.

The DeepMerge network, first introduced by~\cite{CS2020}, is a simple CNN comprised of three convolutional layers followed by batch normalization, max pooling, and dropout, and three dense layers. 
In this paper, the dropout layers have been removed such that the only regularization happens via L2 regularization of the weight decay parameter in the optimizer.
Additionally, the last layer of the original DeepMerge network was updated to include two neurons rather than one. For more details about the architecture check Table~\ref{table:arch} in the Appendix.  

ResNets were first proposed in the seminal paper ~\citet{HZ2015} and have become one of the most widely-used network architectures for image recognition. They are comprised of residual blocks; in the case of ResNet18, blocks of two 3x3 convolutional layers are followed by a ReLU nonlinearity. The chaining of these residual blocks enables the network to retain high training accuracy performance even with increasing network depth.

The domain classifier used in adversarial domain training to calculate transfer loss comprises of three dense layers, the first of which is the same dimension of the extracted features in the base network (either DeepMerge or ResNet18), such that these features form the input into the domain classifier. The second layer has 1024 neurons, followed by ReLU activation and dropout of $0.5$, and the third has one output neuron followed by Sigmoid activation, conveying the domain chosen by the network.

Details about training the networks can be found in Appendix~\ref{sec:networks_params}. We also list all hyperparameters used for training in different experiments in Table~\ref{table:params} and Table~\ref{table:params2}. Our parameter choice for each experiment was informed by running hyperparameter searches using DeepHyper~\citep{BS2018,BE2019}.

\section{Data}
\label{sec:data}

Here we present two dataset pairs for classifying distant merging galaxies ($z=2$) and nearby merging galaxies  ($z<0.1$).

\subsection{Simulation-to-Simulation: Distant Merging Galaxies from Illustris}

It is often very difficult to obtain real-sky observational data of labeled mergers for deep learning models, especially at higher redshifts. Therefore, both the source and target domain of our distant merging galaxy dataset are simulated.

We use the same dataset as in ~\cite{CS2020}, where the authors extract galaxies at redshift $z=2$ from Illustris-1 cosmological simulation~\citep{VG2014}. The objects in this dataset are labeled as mergers if they underwent a major merger (stellar mass ratios of $10:1$) in a time window of $0.5\,\mathrm{Gyr}$ around when the Illustris snapshot was taken. This means our merger sample includes both past (happened before the snapshot) and future mergers (happened after the snapshot). In~\cite{CS2020}, images contain two filters, which mimic Hubble Space Telescope (HST) observations. In this paper we add a third (middle) filter to produce three filter HST images (\textit{ACS F814W, NC F356W, WFC3 F160W}). This allows us to use the images even with a more complex ResNet18 architecture.

We produce two groups of images: source "pristine" images are convolved with a point-spread function (PSF); and target "noisy" images are convolved with the PSF, with added random sky shot noise. This sky shot noise produces a $5\sigma$ limiting surface brightness of 25 magnitudes per square arc-second. More details about the dataset can be found in~\cite{CS2020} and~\cite{SR2019}. The source and target domains contain $8120$ mergers and $7306$ non-mergers, respectively. Each image is $75\times 75$ pixels. We divide these datasets into training, validation, and testing samples: $70\%:10\%:20\%$.
 
\begin{figure}
	\includegraphics[width=0.99\columnwidth]{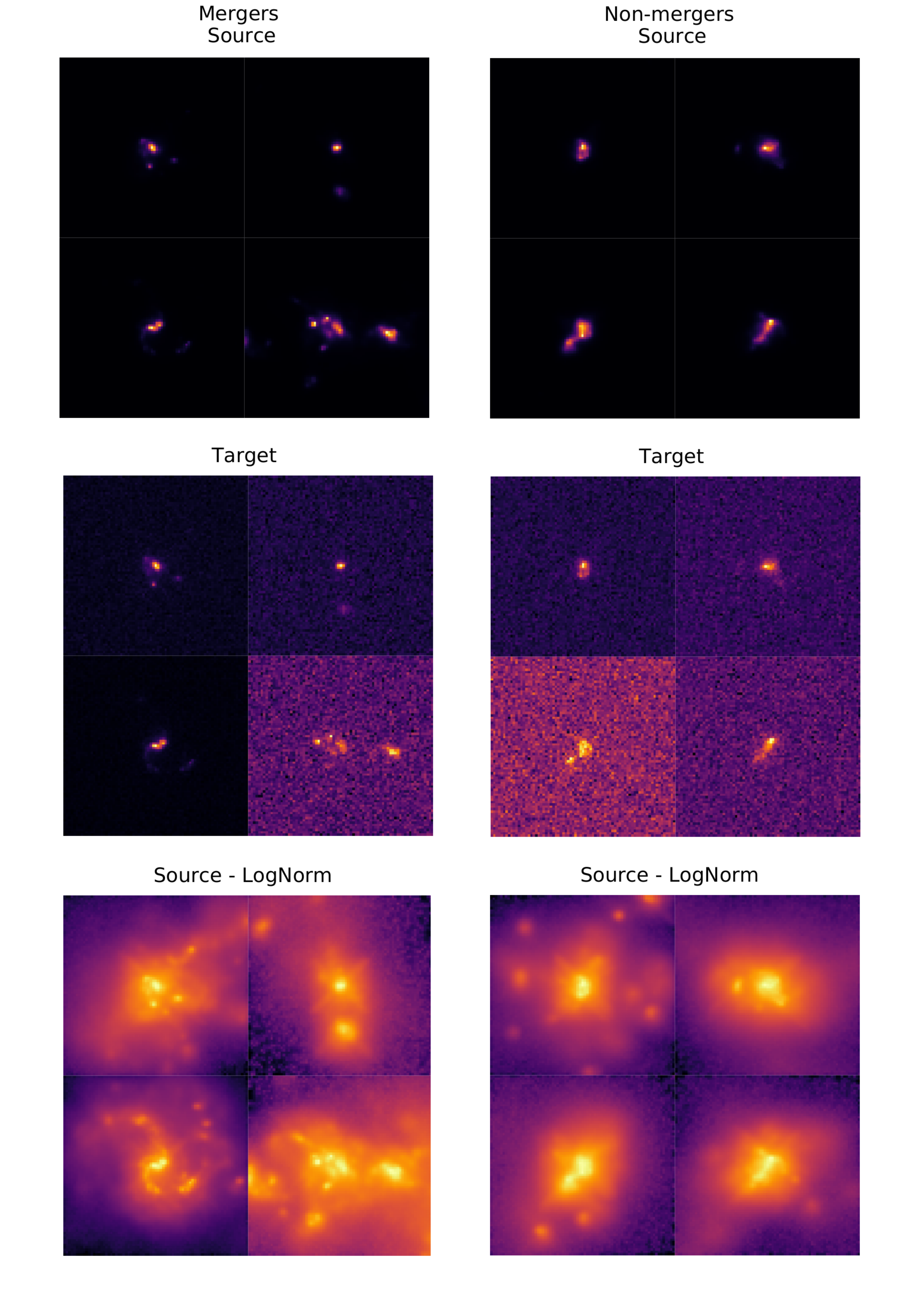}\\
    \caption{Galaxy images from Illustris-1 simulation at $z=2$. The left column shows merging galaxies and the right column shows non-mergers. The same objects are repeated across rows, with the top showing the source domain, the middle showing the target domain, and the bottom displaying the source objects with logarithmic color map normalization for enhanced visibility.}
    \label{fig:examples1}
\end{figure}

See Figure~\ref{fig:examples1} for example images from this Illustris dataset: mergers are shown in the left column and non-mergers in the right column. The top row shows images from the source domain, while the middle row shows the target galaxies with the added noise. The bottom row shows the same group of top-row source images with logarithmic color mapping in order to make the galaxies more visible to the human eye.
 
\subsection{Simulation-to-Real: Nearby Merging Galaxies from Illustris and Real SDSS Images}
\label{sec:simreal}

To test the ability of MMD and adversarial training in the astronomical situations where they show the most promise--- training on simulated data with the prospect of applying the models to real data--- we need to use merging galaxies at lower redshifts where more real data is available. 

\subsubsection{Source dataset: simulated images}

\indent In this scenario, attempting to make simulated images resemble real observations from a particular telescope is an important step for domain adaptation techniques, since decreasing differences between domains will make domain adaptation easier. Here our source data is comprised of simulated merging galaxies (major mergers) from the final snapshot at $z=0$ of Illustris-1, in a time window of $0.25\,\mathrm{Gyr}$ before the time the snapshot was taken. The fact that we use the final snapshot of the simulation is an extremely important difference between this source domain dataset and the one described earlier: this means that only past mergers (also called post-mergers) are included instead of both past and future mergers. Since the simulation was stopped after this snapshot, images of future mergers that would have merged during $0.25\,\mathrm{Gyr}$ after the snapshot are not available. \\
\indent This dataset was originally produced in~\cite{ST2015}. Here, images also include effects of dust implemented as a slab model based on the gas and metal density along the line of sight to each pixel, similar to models by ~\citet{NP2018},~\citet{DL2007}, and~\citet{KW2007}. Images have three SDSS filters ($g$,~$r$,~$i$) and are also convolved with a Gaussian PSF ($\mathrm{FWHM}=1\,\mathrm{kpc}$) and re-binned to a constant pixel scale of $0.24\,\mathrm{kpc}$. This scale corresponds to $1\,\mathrm{arcsec}$ seeing for an object observed by SDSS at $z=0.05$. Finally, random sky shot noise was added to these images by independently drawing from a Gaussian distribution for each pixel, which produces average signal-to-noise ratio of $25$. \\
\indent The simulated galaxies in this dataset contain a lower number of mergers compared to the $z=2$ snapshot used in our simulation-to-simulation experiments. Observational evidence shows that merger rates today are much lower compared to the merger rate peak during the "cosmic high noon" at $z\sim 2-3$~\citep{MP2014}, which is where our galaxies from the previous example are located. Our source domain dataset contains only $44$ post-mergers and $5625$ non-mergers. We employ data augmentation in order to make a larger source dataset, particularly focusing on mergers to make the classes balanced. We first augment mergers by using mirroring (vertical and horizontal), and rotation by $90^{\circ}$ and $180^{\circ}$ (which produces $220$ images). Finally, these images are additionally augmented by random angle rotation or zooming in/out. The final source dataset we use contains $3000:3000$ post-mergers and non-mergers (we truncate non-mergers to make the classes balanced).

\subsubsection{Target dataset: observed images}

\indent Our target dataset is composed of observational SDSS images. We follow dataset selection from~\cite{AS2018} and use the SDSS online image cutout server to get RGB (red, green, blue) JPEG images of both merging and non-merging galaxies. These RGB images correspond to ($i$,~$r$,~$g$) SDSS filters, as opposed to ($g$,~$r$,~$i$) in our source domain. We later align the order of filters in our source domain to correspond to this filter order. All of the galaxies selected are from the Galaxy Zoo project~\citep{LS2008,LS2010}, which used crowd-sourcing to generate labels for 900,000 galaxies. We use the 3003 mergers identified in the~\cite{DK2010} catalogue; three of these mergers were unable to be retrieved due to faulty weblinks. This catalogue identified mergers through the weighted-merger-vote fraction, $f_{m}$, which describes the confidence in the crowd-sourced label. Mergers were defined as galaxies with an $\ f_{m}>0.4$, where $\ f_{m}=0$ describes objects that are not merger-like and $\ f_{m}=1$ describes merger-like objects. Mergers included in Galaxy Zoo were also required to be between redshifts $0.005 < z < 0.1$. \\
\indent All available SDSS mergers also include a merger stage: separated mergers ($167$ images), interacting ($2523$ images), post-mergers ($310$ images). Since our source domain includes only post-mergers, we restrict our target dataset to only include the post-merger subclass from SDSS. To obtain $3000$ mergers, we augment the SDSS post-merger images using the same techniques used in the source domain. To complete our target dataset, another $3000$ non-merger galaxies in the $0.005 < z < 0.1$ redshift range were randomly selected from the Galaxy Zoo project’s entire dataset by requiring $\ f_{m}<0.2$.

\begin{figure}
	\includegraphics[width=0.99\columnwidth]{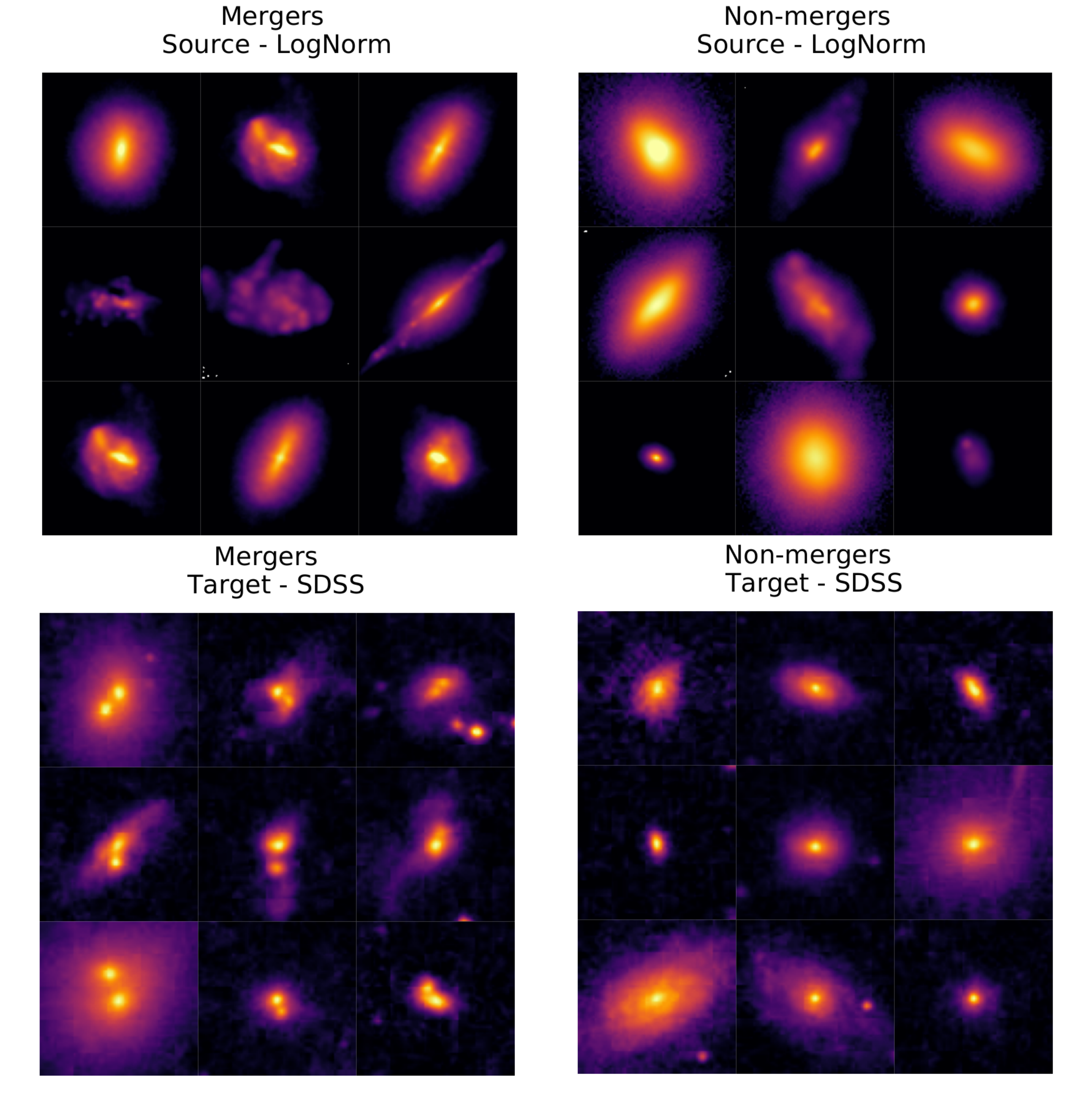}\\
    \caption{Galaxy images from Illustris simulation at $z=0$ mimicking SDSS observations (top row) and real SDSS images (bottom row). The left column shows post-merger galaxies, while the right column shows non-mergers. Source domain images in the top row were plotted with a logarithmic color map to make features more visible. Even when we select only post-mergers from SDSS we can still see that the merger class is different across the two domains. While the source domain contains more relaxed systems, the target contains galaxies near each other, with two bright clearly visible cores.}
    \label{fig:examples2}
\end{figure}

We resize images from both domains to the same size as in simulation-to-simulation example ($75\times75$ pixels), and use the same split into training, validation and testing samples: $70\%:10\%:20\%$. In Figure~\ref{fig:examples2} we plot images from both domains. In the top row, we plot post-mergers (left) and non-mergers (right) from Illustris simulation at $z=0$, while in the bottom row we plot post-mergers (left) and non-mergers (right) from SDSS.

Images from the target domain were visually classified, and most of the target post-mergers clearly exhibit two bright galaxy cores, while images from the source domain display a greater variety of characteristics. Consequently, the two domains are extremely dissimilar relative to the two domains in the simulation-to-simulation example. The choices we detail above--- using only post-mergers in both domains; including observational and dust effects and choosing a small time window to avoid the inclusion of very relaxed merger systems in the source domain--- were made in an attempt to make the two domains as similar as possible. Still, the fact that the number of individual mergers in both domains is very small, paired with the fact that their appearance can be quite different makes any domain adaptation efforts quite challenging.

\section{Experiments}
\label{sec:experiments}

CNNs outperform other machine learning methods for classification of merging galaxies~\citep{SR2019,CS2020}. However, in both~\cite{CS2020} and~\cite{PW2019}, it was shown that, even though training and evaluating a CNN on images from the same domain gives very good results, a simple model trained in one domain cannot perform classification in a different domain with high accuracy. To increase the performance of deep learning classifiers on a target dataset, we use the DA techniques described in Section~\ref{sec:methods}. 

We first train neural networks without the implementation of any DA techniques to determine the base performance for source and target images for each pair of datasets: simulated-to-simulated (Illustris $z=2$ with and without noise) and simulated-to-real (Illustris $z=0$ and SDSS). While we possess labels for both the source and target domain in both scenarios, this training is performed using labeled source images exclusively. We only use the target image labels in the testing phase to asses accuracy. We seek these performance accuracies as the metric to improve upon through domain adaptation. 

We then run several domain adaptation experiments--- using MMD as transfer loss, adversarial training with DANN domain discriminator loss as transfer loss, MMD as transfer loss with Fisher loss and entropy minimization, and finally DANN adversarial training with Fisher loss and entropy minimization--- with both DeepMerge and ResNet18 architectures on the simulation-to-simulation dataset. Training with domain adaptation is performed using labeled source data and unlabeled target data. In the case of adversarial training, an additional domain classifier branch is added to receive an input of features from  base network. The parameters used for training both DeepMerge and ResNet18 for all simulation-to-simulation experiments are given in the Appendix in Table~\ref{table:params}.

Since larger networks are prone to overfitting, given the limited size of both source and target datasets in our simulation-to-real experiments, we decided to only test it with the the smaller DeepMerge network. Similar to the experiments described above for the simulation-to-simulation dataset, we first train DeepMerge without any domain adaptation in order to determine the base performance. We then tried to improve target domain accuracy by training using MMD and adversarial training, both with and without Fisher loss and entropy minimization. However, despite performing hyperparameter searches, domain adaptation was not successful, i.e. the target domain accuracy was no better than random guessing. We then turned our trials to combining MMD and adversarial training with transfer learning from models successfully trained in simulation-to-simulation experiments. Hyperparameters used in training the model in the simulation-to-real experiments are given in the Appendix in Table~\ref{table:params2}.

To ensure reproducibility of our results prior to training, we fix the random seeds used for image shuffling (before division into training, validation and testing samples), as well as for random weight initialization of our neural networks. The same images were used across experiments for training, as well as for testing afterwards to produce the reported results. In Section~\ref{sec:results} we report results for a fixed seed=1.

\section{Results}
\label{sec:results} 

Throughout this paper we consider mergers the positive class (label $1$), and non-mergers the negative class (label $0$). Consequently, correctly/incorrectly classified merger are true positives (TP)/false negatives (FN), while correctly/incorrectly classified non-mergers are true negatives (TN)/false positives (FP). \\
\indent We report classification accuracy, precision or purity: TP/(TP + FP), recall or completeness: TP/(TP + FN), and $\mathrm{F1\,score} = 2\frac{\mathrm{Precision}\times\mathrm{Recall}}{\mathrm{Precision}+\mathrm{Recall}}$. We also report the Area Under the Curve (AUC) score--- the area under the Receiver Operating Characteristic (ROC) curve, which conveys the trade-off between true-positive rate and false-positive rate. Finally, we provide Brier score values, which measure the mean squared error between the predicted scores and the true labels; a perfect classifier would have a Brier score of zero. 

\subsection{Simulation-to-Simulation Experiments}
\label{sec:results_ss}

\begin{table*}
   \centering
   \noindent\begin{minipage}[b]{0.99\textwidth}
   \centering
    \caption{Performance metrics of the DeepMerge and ResNet18 CNNs, on source and target domain test sets, without domain adaptation (first row) and when domain adaptation techniques are used during training (all other rows).
    The table shows AUC, Accuracy, Precision, Recall, F1 score, and Brier score.}
  \label{table:performance}
  \centering
  \begin{tabular}{|l | l | c c| c c|}
    \multicolumn{2}{c}{}  & \multicolumn{4}{c}{Simulated-to-Simulated} \\\hline 
\multirow{2}{*}{Loss}      &  \multirow{2}{*}{Metric}     &  \multicolumn{2}{c|}{Source}   &   \multicolumn{2}{c|}{Target}  \\
                                                &               & DeepMerge       & ResNet18 & DeepMerge       & ResNet18 \\\Cline{1.2pt}{1-6}
\multirow{5}{*}{No Domain Adaptation}           &  AUC          &   $0.92$        &   $0.88$ &     $0.74$      &  $0.73$    \\
                                                &  Accuracy     &   $0.85$        & $0.81$   & $0.58$          & $0.60$    \\ 
                                                &  Precision    &   $0.88$        &   $0.82$ &      $0.99$     &  $0.82$   \\
                                                &  Recall       &   $0.83$        &  $0.83$  &      $0.08$     &  $0.31$    \\
                                                &  F1 score     &   $0.86$        &  $0.83$  &     $0.14$      &  $0.45$    \\
                                                &  Brier score  &   $0.11$        &  $0.14$  &      $0.47$     &  $0.23$   \\\hline
\multirow{5}{*}{MMD}                            &  AUC          &   $0.93$        &  $0.96$  &     $0.85$      &  $0.80$    \\
                                                &  Accuracy     & $0.87 $         & $0.90$   &      $0.77$     & $0.74$    \\ 
                                                &  Precision    &   $0.88$        &  $0.93$  &    $0.81$       &  $0.75$    \\
                                                &  Recall       &     $0.87$      &  $0.90$  &     $0.72$      &  $0.74$    \\
                                                &  F1 score     &    $0.88$       &  $0.91$  &     $0.76$      &  $0.75$     \\              
                                                &  Brier score  &   $0.10$        &  $0.08$  &      $0.17$     &  $0.21$     \\\hline
\multirow{5}{*}{MMD + Fisher + Entropy}         &  AUC          &    $0.92$       &  $0.94$  &    $0.86$       & $0.81$     \\
                                                &  Accuracy     &   $0.84$        & $0.89$   &   $0.77$        & $0.75$      \\ 
                                                &  Precision    &    $0.87$       &   $0.90$ &    $0.79$       &  $0.75$   \\
                                                &  Recall       &    $0.84$       &  $0.90$  &     $0.75$      &  $0.78$    \\
                                                &  F1 score     &    $0.86$       &  $0.86$  &     $0.77$      &  $0.77$   \\
                                                &  Brier score  &   $0.11$        &   $0.09$ &      $0.16$     &  $0.19$    \\\hline
\multirow{5}{*}{Adversarial}                    &  AUC          &   $0.94$        &   $0.97$ &  $0.87$         & $0.78$     \\
                                                &  Accuracy     & $0.87$          & $0.92$   & $0.79$          & $0.72$      \\ 
                                                &  Precision    &    $0.88$       &  $0.93$  &   $0.79$        &  $0.74$   \\
                                                &  Recall       &    $0.89$       &  $0.93$  &   $0.81$        &  $0.71$     \\
                                                &  F1 score     &    $0.87$       &   $0.93$ &    $0.80$       & $0.72$      \\
                                                &  Brier score  &   $0.09$        &  $0.06$  &      $0.16$     & $0.25$     \\\hline
\multirow{5}{*}{Adv. + Fisher + Entropy}        &  AUC          &     $0.94$      &  $0.94$  &    $0.82$       &  $0.75$  \\
                                                &  Accuracy     &     $0.87$      & $0.87$   &   $0.74$        & $0.70$   \\ 
                                                &  Precision    &   $0.90$        &  $0.92$  &   $0.78$        & $0.83$     \\
                                                &  Recall       &    $0.86$       &  $0.85$  &     $0.68$      &  $0.54$   \\
                                                &  F1 score     &   $0.88$        &  $0.88$  &    $0.73$       &  $0.65$    \\
                                                &  Brier score  &   $0.10$        &  $0.09$  &      $0.21$     &  $0.24$  \\\hline
\end{tabular}
\end{minipage}
\end{table*}

\begin{figure*}
	\includegraphics[width=1.24\columnwidth]{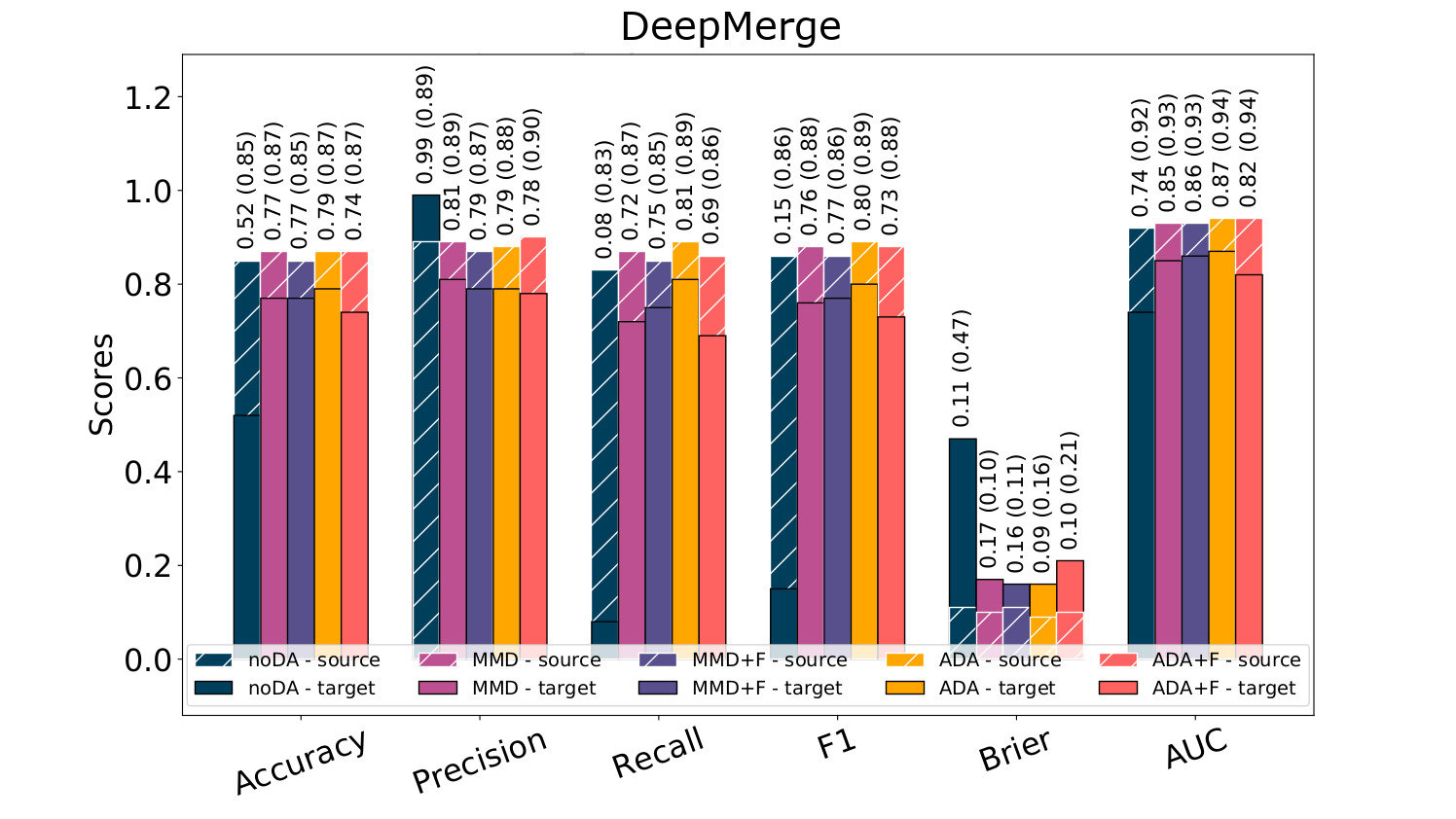}
	\includegraphics[width=.74\columnwidth]{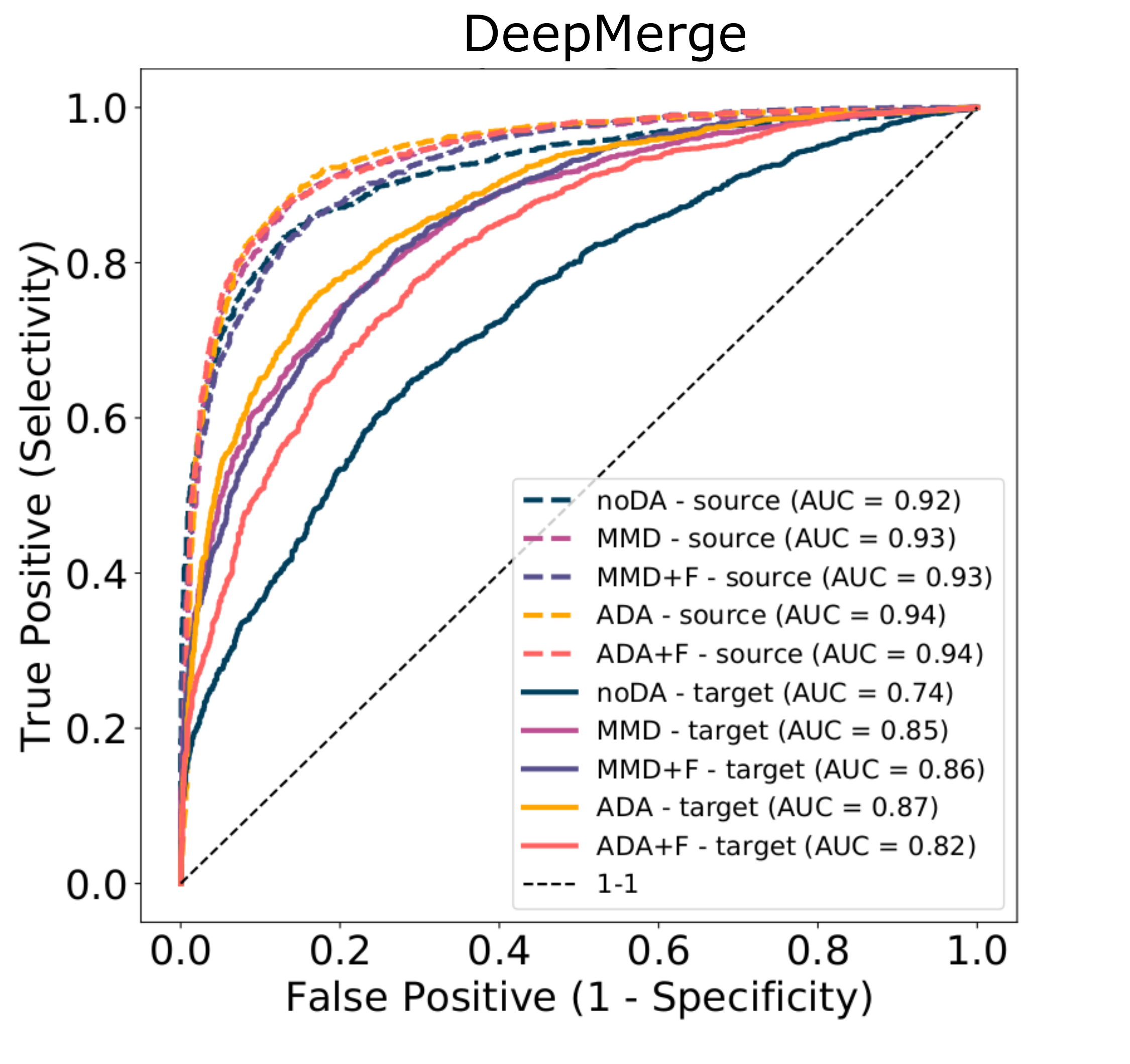}\\\hfill
	\includegraphics[width=1.24\columnwidth]{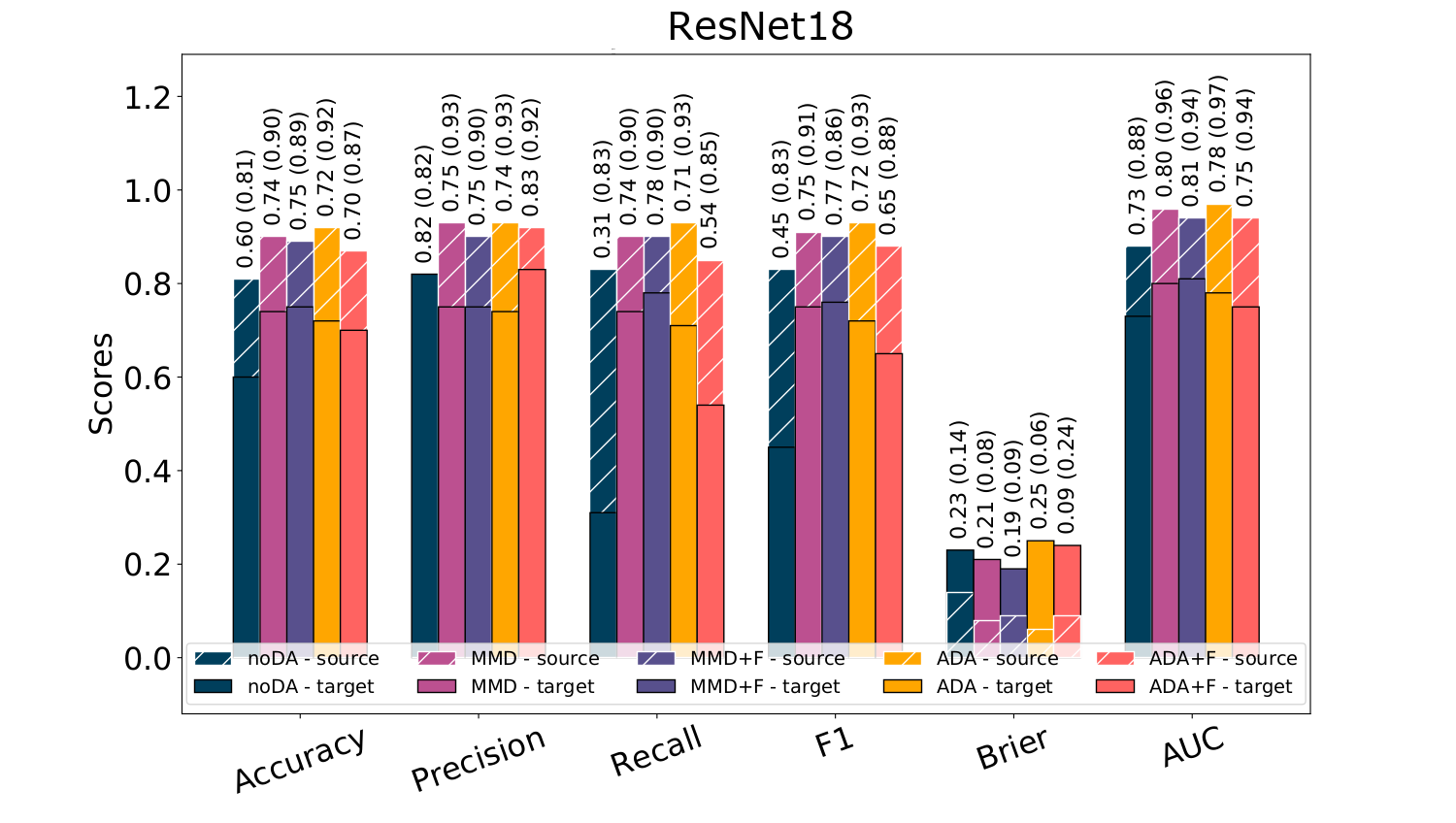}
	\includegraphics[width=.74\columnwidth]{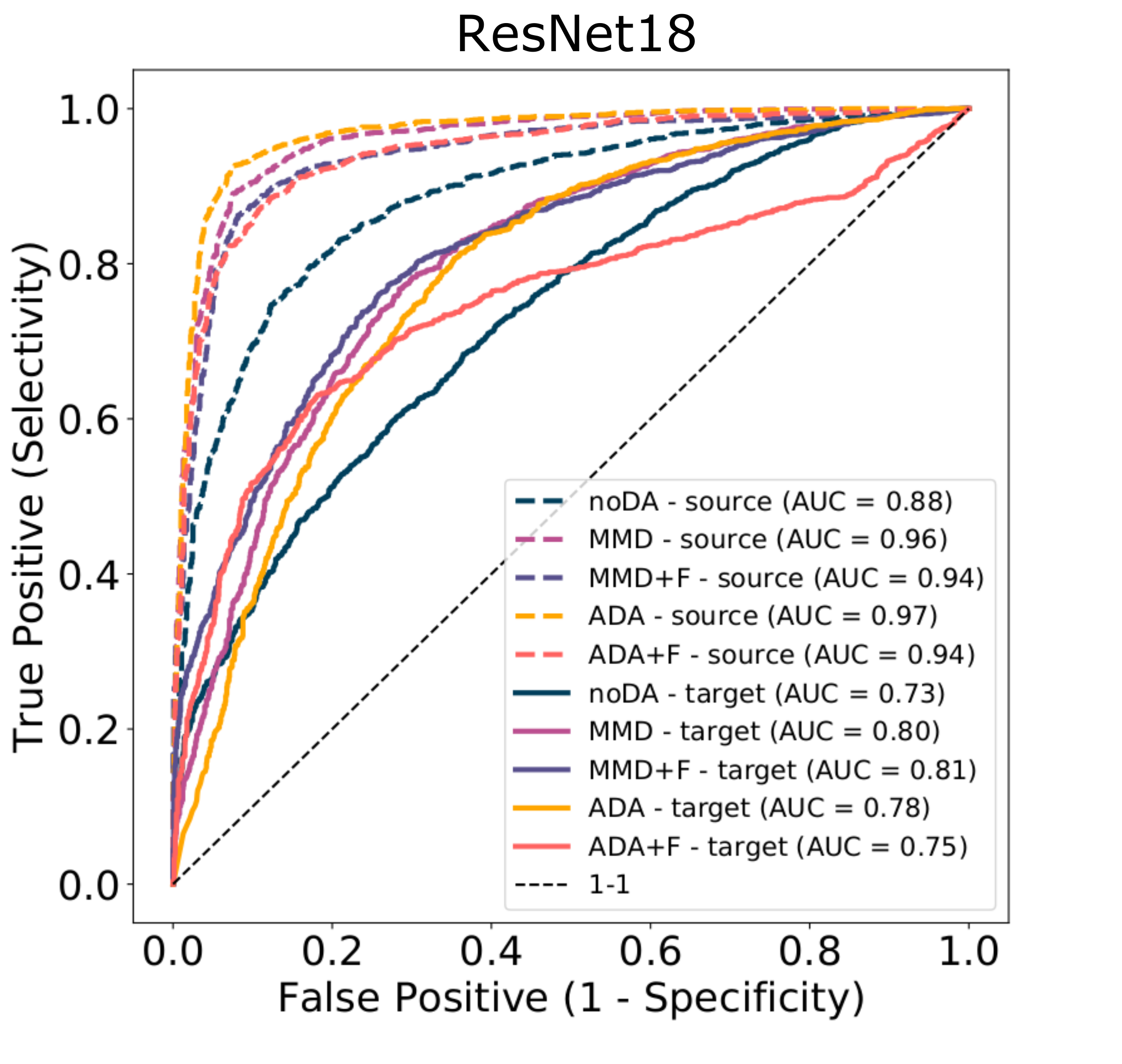}\hfill
    \caption{The top panel shows classification results for DeepMerge network and the bottom panel for ResNet18. Left: Performance metrics for no domain adaptation experiment (labeled "noDA") in navy blue, MMD in purple, MMD wih Fisher loss and entropy minimization (labeled "MMD+F") in dark purple, adversarial training (labeled "ADA") in yellow and adversarial training with Fisher loss and entropy minimization (labeled "ADA+F") in pink. We plot values for accuracy, precision, recall, F1 score, Brier score, and AUC. Dashed bars show results for the source domain and solid colored bars for the target domain. Right: ROC curves with the same color and line style scheme. In the legend we also give AUC values for all five experiments.}
    \label{fig:bar}
\end{figure*}

Results of training the two classifiers without domain adaptation are given in the first row of Table~\ref{table:performance}. Training was performed on the source domain images, and test accuracy on source images is high for both networks in the base case without DA: $85\%$ for DeepMerge and $81\%$ for ResNet18. As was expected, without any domain adaptation both classifiers are almost unable to classify target domain images; test accuracy for this domain are only $58\%$ (DeepMerge) and $60\%$ (ResNet18). Additionally, as expected, we noticed that the more complex ResNet18 was much more prone to overfitting earlier in the training than DeepMerge. We therefore implemented early stopping in our training in all experiments, as well as saving of the best model before the training stops due to substantial overfitting.

We then trained DeepMerge and ResNet18 with both domain adaptation techniques, each with and without Fisher loss and entropy minimization. Results from these DA experiments are also given in Table~\ref{table:performance}. We conclude that it is difficult to determine a single best technique across architectures. Additionally, inclusion of the Fisher loss and entropy minimization, implemented for within class compactness in both the source and target domains, does not always help. This might be due to the fact that multiple losses interact differently depending on the network architecture and complexity of the feature space. In short, simple hyperparameter grid searches, which informed our parameter choices, are not a perfect solution to find the optimal hyperparameters for different experiments (a non-trivial task that we leave for future studies). Despite our imperfect hyperparamter choices, we assert that the results presented here convey an overall demonstration of the performance and improvements of domain adaptation techniques for cross-domain studies.\\
\indent Next we take a closer look at experiments that were most successful for DeepMerge and ResNet18. The best-performing DeepMerge network experiments--- MMD, adversarial training, and adversarial training with Fisher loss and entropy minimization--- reached source domain accuracies of $87\%$. The accuracy in the target domain was largest with adversarial training at $79\%$, while with MMD it reached $77\%$. Consequently, the highest increase in target domain accuracy was $21\%$ compared to the classifier without domain adaptation. Again, we assert that each experiment's results could potentially be further improved with a different set of hyperparameters.\\
\indent For ResNet18, we see a slightly smaller increase in target domain accuracies compared to improvements by DeepMerge. Because it is a more complex network, it is harder to stop ResNet18 from learning more intricate details that are only found in the source domain. Target domain accuracies increase from $60\%$ without domain adaptation to $75\%$ in the best performing experiment, which was MMD with additional Fisher loss and entropy minimization. \\
\indent In experiments without DA, we allow the network to learn from all available features that can be extracted rather than restricting to the set of domain-invariant features. It follows that, in training without DA, a perfectly optimized network should be able to reach its highest source accuracy when not being forced to learn domain-invariant features. In contrast to this expectation, we observe that the addition of transfer and other losses slightly increases source domain accuracies in almost all of the experiments we ran. We posit that this is first and foremost the consequence of not finding the best set of hyperparameters and that the additional transfer loss serves as a good regularizer, enabling longer training without overfitting. This is more prominent in case of ResNet18, where source domain accuracies increase from $81\%$ for no domain adaptation case to the highest value of $92\%$ in the case of adversarial training. \\
\indent For easier comparison, in Figure~\ref{fig:bar} we plot the performance values from Table~\ref{table:performance} for our test set of images. The top row of plots shows results for the DeepMerge network, while the bottom row is for ResNet18. Bar plots on the left show all performance metrics (accuracy, precision, recall, F1 score, Brier score, and AUC) for our simulated-to-simulated experiments: no domain adaptation (navy blue), MMD (purple), MMD with Fisher and entropy minimization (dark purple), adversarial training (yellow), and adversarial training with Fisher and entropy (pink). The two right panels show ROC curves for all DeepMerge and ResNet18 experiments, with the same color coding as in the bar plots. In all four panels, the solid bars and lines show values for the target domain while the dashed bars and lines show source domain performance. 

See Appendix~\ref{sec:networks_results} for additional performance comparison between training with and without domain adaptation for both networks on the simulation-to-simulation dataset.

\subsection{Simulation-to-Real Experiments}
\label{sec:results_sr}

We also evaluated the performance of these DA methods in the situation where the classifier is trained on simulated source domain images and tested on a target domain of observational telescope images. Examples like this are much more complex than our simulation-to-simulation experiments due to the larger discrepancy between domains, as discussed at length in Subsection~\ref{sec:simreal}.

Due to the perils of training a large network on such a small dataset, we decided to test domain adaptation techniques with only the smaller DeepMerge network. Without DA, DeepMerge reached an accuracy of $92\%$ in the source domain and $50\%$ in the target domain in the testing phase. This was our baseline we try to improve upon in the simulation-to-real DA experiments. Due to the extreme discrepancy between domains, we also tested trained the DeepMerge network on the target domain directly, which resulted in an accuracy of $96\%$ on the target domain images. This is higher than for the source domain of simulation images, confirming that the target domain is easier to train on since it is comprised of visually more apparent merging features.\\
\indent We then tried running hyperparameter grid searches with MMD and adversarial training, but were unable to successfully use domain adaptation to improve target domain accuracies. This led us to conclude that problems with the size of our dataset, as well as the difference between domains, was preventing the successful learning of domain-invariant features. In the top row of Table~\ref{table:performance2}, we report all performance metrics for no domain adaptation case and in the middle row we report results of using MMD only as transfer loss. Since no other method was successful, we omit reporting numbers for all other DA setups tested in the simulation-to-real experiments. Larger simulated training samples, more sophisticated domain adaptation methods that allow for better domain overlap of discrepant feature distributions, or a combination of the two will advance the study of merging galaxies across the simulated-to-real domain in the future.

\begin{table}
   \centering
   \noindent\begin{minipage}[b]{0.99\columnwidth}
   \centering
    \caption{Performance metrics of DeepMerge, on source simulated data and target observational data in the testing phase: without domain adaptation (first row), MMD only (middle row), and MMD with tranfer learning (bottom row).
    The table shows AUC, accuracy, precision, recall, F1 score, and Brier score.}
  \label{table:performance2}
  \centering
  \begin{tabular}{|l | l |c c|}
    \multicolumn{1}{c}{}  &  \multicolumn{3}{c}{Simulated-to-Real} \\\hline
 \multirow{2}{*}{Loss}       &   \multirow{2}{*}{Metric}   &  \multicolumn{2}{c|}{DeepMerge} \\
     &          & Source  &  Target \\\Cline{1.2pt}{1-4}
\multirow{5}{*}{No Domain Adaptation}           &  AUC          &  $0.97$        &  $0.58$      \\
                                                &  Accuracy     &   $0.92$       &  $0.50$    \\ 
                                                &  Precision    &   $0.91$       &  $0.50$   \\
                                                &  Recall       &   $0.92$       &  $0.80$   \\
                                                &  F1 score     &    $0.92$      &  $0.62$   \\
                                                &  Brier score  &    $0.06$      &  $0.49$   \\\hline
\multirow{5}{*}{MMD}                            &  AUC          &  $0.98$        &  $0.60$   \\
                                                &  Accuracy     &   $0.94$       &  $0.53$   \\ 
                                                &  Precision    &   $0.92$       &   $0.53$   \\
                                                &  Recall       &   $0.95$       &   $0.63$     \\
                                                &  F1 score     &   $0.94$       &   $0.58$  \\              
                                                &  Brier score  &    $0.05$      &   $0.40$   \\\hline
\multirow{5}{*}{Transfer Learning + MMD}         &  AUC          &    $0.90$      &  $0.76$     \\
                                                &  Accuracy     &    $0.83$      &   $0.69$   \\ 
                                                &  Precision    &    $0.89$      &   $0.68$     \\
                                                &  Recall       &    $0.74$      &   $0.74$     \\
                                                &  F1 score     &   $0.80$       &  $0.71$  \\            
                                                &  Brier score  &   $0.13$       &  $0.23$  \\\hline
\end{tabular}
\end{minipage}
\end{table}

\begin{figure*}
	\includegraphics[width=1.02\columnwidth]{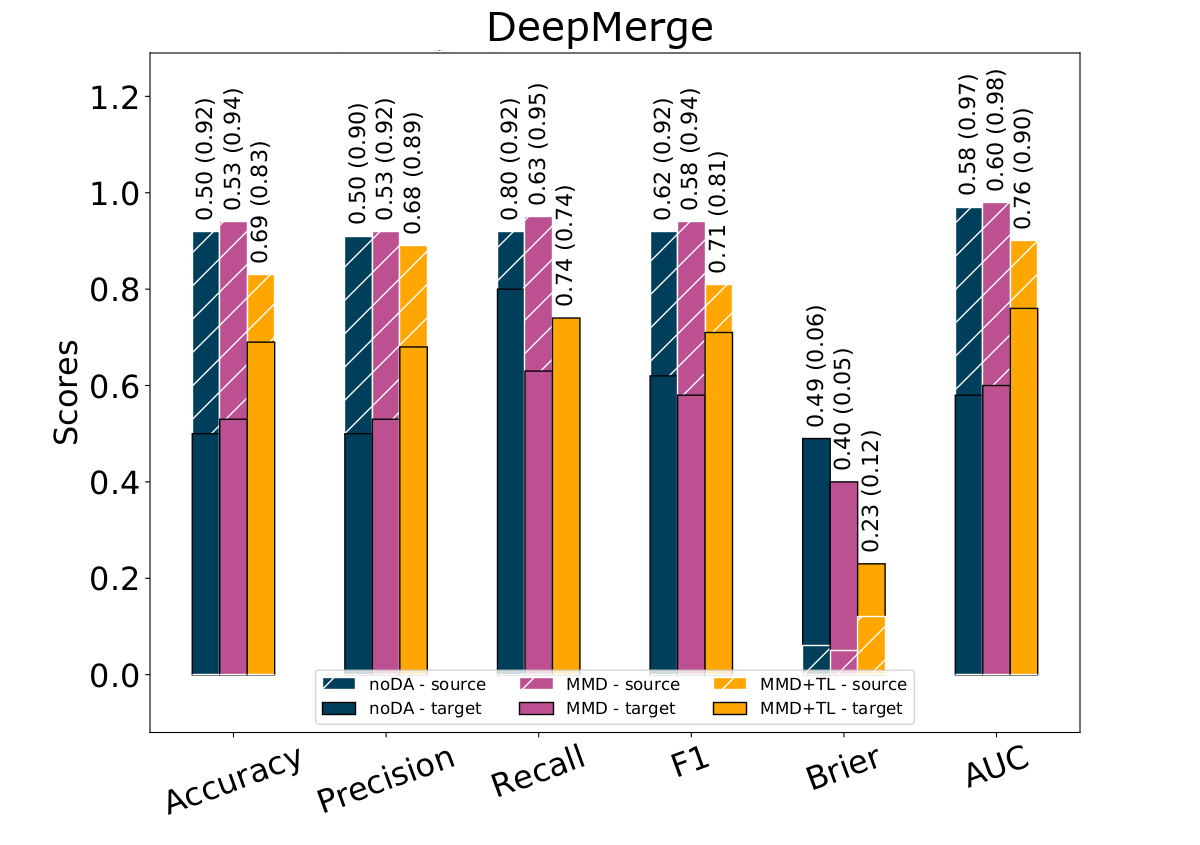}
	\includegraphics[width=.76\columnwidth]{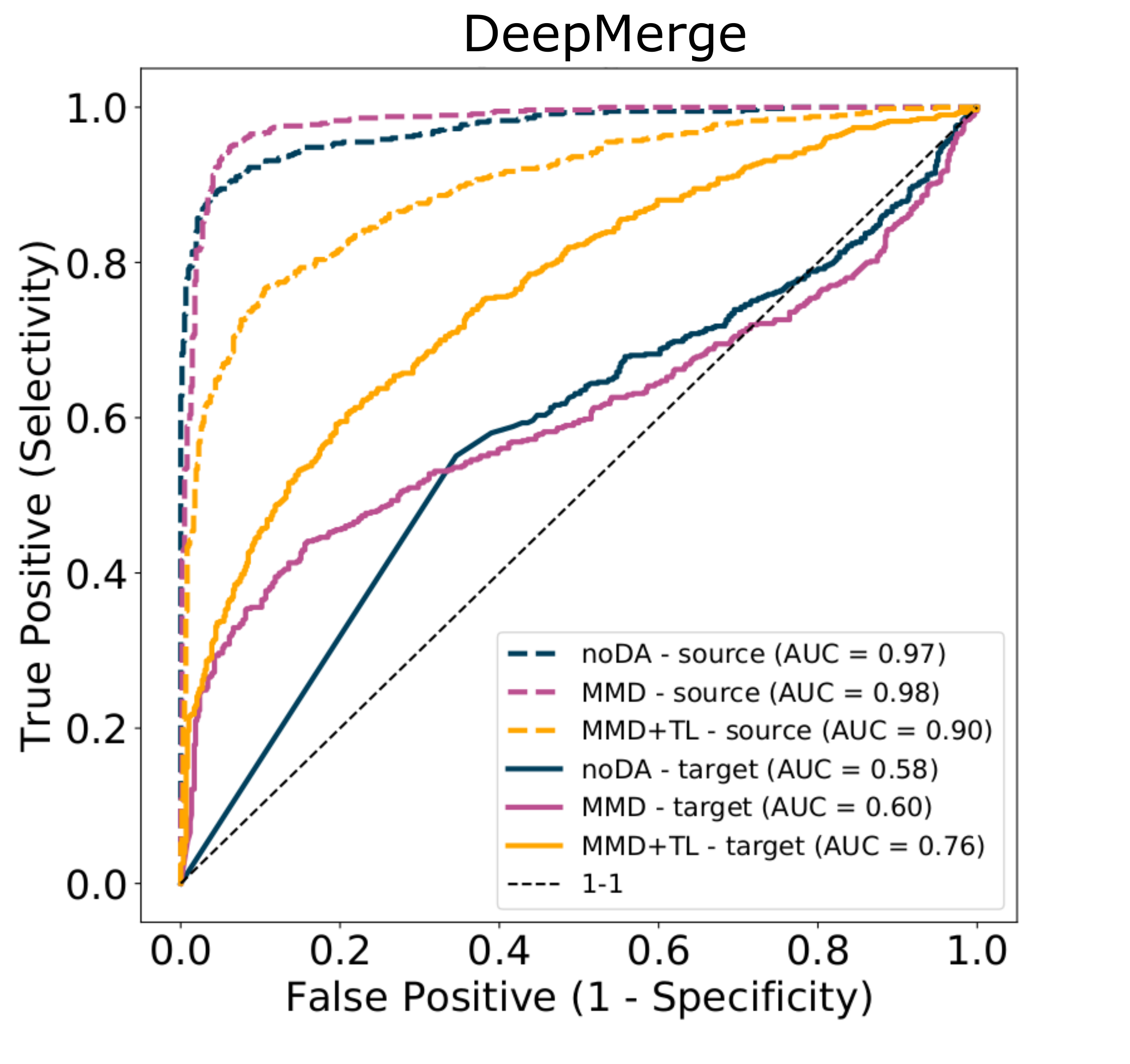}\\\hfill
    \caption{Left: Performance metrics for DeepMerge network for no domain adaptation experiment (labeled "noDA") in navy blue, MMD in purple, MMD with transfer learning (labeled "MMD+TL") in yellow. We plot values for accuracy, precision, recall, F1 score, Brier score, and AUC. Dashed bars show results for the source domain and solid colored bars for the target domain. Right: ROC curves with same color and line style scheme.  In legend we also give AUC values for all three experiments.}
    \label{fig:bar2}
\end{figure*}

\subsubsection{Transfer Learning}
The approach we took to overcome our small dataset limitation was to use transfer learning, where the weights from a neural network pre-trained on different data are loaded before training the classifier on the data of interest. 

Transfer learning has been used in previous studies of merging galaxies. For example, in~\cite{AS2018}, authors use Xception~\citep{CH2016}, a large deep learning model, pre-trained on images of everyday objects from ImageNet~\citep{DD2009}. They successfully trained the model on observed images of merging galaxies from SDSS and report classification precision, recall, and F1 score of $0.97,0.96,0.97$. Similarly, in~\cite{WP2020}, authors use a VGG network~\citep{SZ2015} pre-trained on ImageNet to train on simulated images from the IllustrisTNG simulation at $z=0.15$~\citep{SP2018, PS2018}. They report classification accuracy of $72\%$ on simulated images, and then use the simulation-trained model to detect major mergers in KiDS~\citep{JK2013} and GAMA~\citep{DN2009} observations. \\
\indent We decided to test if our simulated-to-real DA setup would benefit from transfer learning from a more similar dataset than ImageNet. Rather than proceeding with random weight initialization, we load the weights from our successfully trained DeepMerge networks in our simulated-to-simulated experiments. This way we can utilize extracted features that relate to distant merging galaxies, which are much more similar to nearby merging galaxies, than features extracted from everyday objects. \\
\indent We tried training without freezing layers (allowing all weights in the network to fine-tune to the new datasets), and freezing of convolutional and batch normalization layers. The training performed much better when all weights of the model were allowed to train from their loaded checkpoint. This may be due to both the smaller size of the DeepMerge network as well as the possible differences in the appearance of galaxies in our experiments, with a particular emphasis on the very different appearance of real SDSS mergers compared to simulated ones. This probably led to the necessity of the network finding better-suited domain invariant features when real data is included, which can be more easily found when convolutional layers are allowed to train. \\
\indent We also performed a hyperparameter search for both MMD and adversarial domain adaptation with transfer learning, and were able to find a configuration for successful DA with MMD. Since the domain discrepancy in the case of simulated-to-real images was large, successfully learning common features led to the reduction of source domain accuracy from $92\%$ without domain adaptation to $83\%$ with MMD and transfer learning. However, and most significantly, we were able to achieve $69\%$ in the target domain during the testing phase of MMD with transfer learning--- an increase of $19\%$ compared to noDA. In the bottom row of Table~\ref{table:performance2}, we report performance metrics for this transfer learning case. \\
\indent For ease of comparison, we also plot performance metric values in the testing phase in Figure~\ref{fig:bar2}. Bar plots on the left show all performance metrics (accuracy, precision, recall, F1 score, Brier score, and AUC) for our simulated-to-real experiments--- no domain adaptation (navy blue), MMD (purple), MMD with transfer learning (yellow). ROC curves for these experiments are presented on the right, with the same color coding as in the bar plots. In both panels, the solid bars and lines show values for the target domain while the dashed bars and lines show source domain performance. \\
\indent This experiment demonstrates that domain adaptation techniques are very powerful. However, to be useful in a scientific context, we conclude that very careful data preprocessing to reduce domain discrepancies and/or transfer learning to mitigate the problem of small datasets is necessary. It is our hope that the introduction of these techniques to the astronomy community will spur innovation and encourage the use of more sophisticated DA methods that optimize domain alignment for this sort of difficult astronomical tasks.

\section{Discussion}
\label{sec:discuss} 

We have demonstrated how MMD and domain adversarial training substantially increase the performance of simulated-to-simulated learning in the context of galaxy merger classification. For DeepMerge, the average accuracy benefit of these techniques was $18.75\%$ in the target domain; for ResNet18, the average benefit was $12.75\%$. While unable to show positive results with adversarial domain adaptation training on our simulated-to-real dataset, pairing MMD with transfer learning achieved a substantial increase in target domain accuracy of $19\%$ with DeepMerge. \\
\indent We believe that both techniques show great promise for use in astronomy. Here we discuss their interpretability with the aid of t-Distributed Stochastic Neighbor Embeddings (t-SNEs) and Gradient-Class Activation Mappings (Grad-CAMs); and provide an outlook on their potential for use within the scientific community.

\subsection{Model Interpretability: Understanding the Extracted Features with t-SNEs}
To better understand the effect of domain adaptation, we visualize the distribution of the extracted features with t-Distributed Stochastic Neighbor Embeddings (t-SNE) plots by projecting the high-dimensional feature space to a more familiar two-dimensional plane~\citep{MH2008}. This method calculates the probability distribution over data point pairs, assigning a higher probability to similar objects and a lower probability to dissimilar pairs, in both the latent feature space and in the two-dimensional mapping.  By minimizing the Kullback–Leibler (KL) divergence~\citep{KL1951} between the two distributions, the t-SNE method ensures the similarity between the actual distribution and the projection. Despite its usefulness, we emphasize that t-SNE is a non-linear algorithm and adapts to  data by performing different transformations in each region. This can lead to clumps with highly-concentrated points appearing as very large groups, i.e. it is difficult to compare relative sizes of clusters in t-SNE plot renderings. Additionally, the outputted two-dimensional embeddings are entirely dependent on several user-defined parameters. For more details on t-SNE best practices, see ~\cite{WV2016}.

\begin{figure*}
	\includegraphics[width=0.95\linewidth]{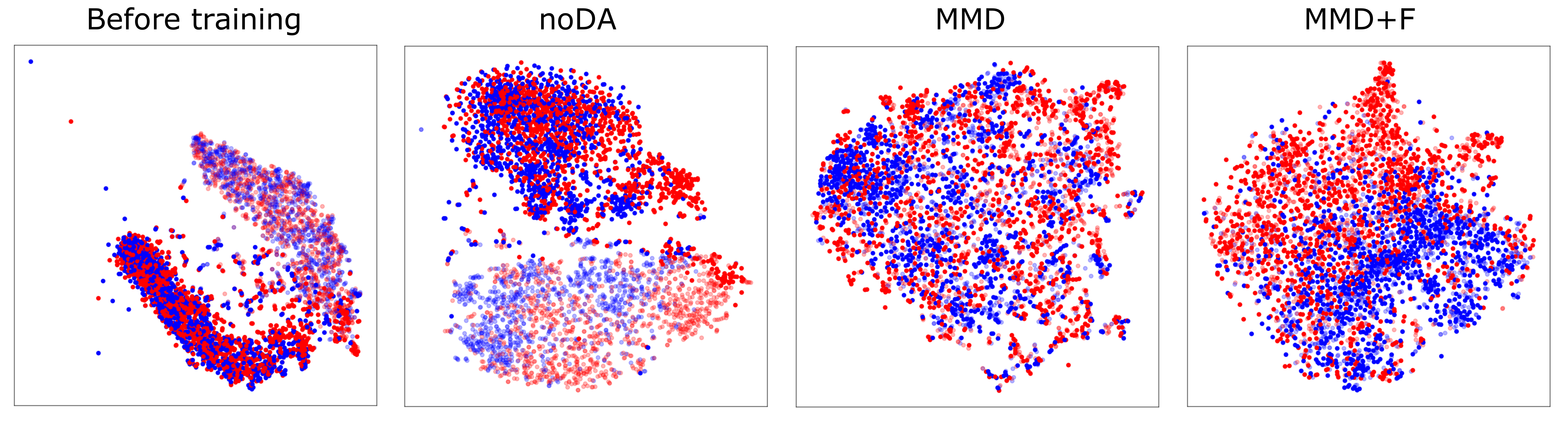}
    \caption{t-SNE plots for the DeepMerge network. Red and blue colored dots represent mergers and non-mergers, respectively; transparent dots represent the source domain, while opaque dots represent the target domain. The first panel shows classes before any training, while all other panels show embeddings after $40$ epochs. The second panel represents training without any domain adaptation, the third panel show training using MMD as transfer loss and the fourth panel shows MMD with Fisher loss and entropy minimization.}
    \label{fig:tsne}
\end{figure*}

We implemented an option to plot t-SNEs in our training method to demonstrate the changes in extracted features from the source and target domain across a series of epochs. In Figure ~\ref{fig:tsne} we plot t-SNE plots for DeepMerge before the start of the training in the first panel, and t-SNE plots after some training--- when no domain adaptation is implemented in the second panel, when MMD as transfer loss is used in the third panel, when MMD with Fisher loss and entropy minimization is used in the fourth panel. We confirm that domain adversarial training t-SNEs are virtually indistinguishable from MMD plots, so we omit the repeat here. On all t-SNE plots, red and blue colored dots represent the two classes--- mergers and non-mergers, respectively--- and transparent dots represent the source domain, while opaque dots represent the target domain.

Before the start of training, classes are completely mixed together and domains are separated (first panel). With no domain adaptation, we see that even after some training, domains remain separated (second panel). With domain adaptation but no Fisher loss or entropy minimization (third panel), we see that features from both domains completely overlap. We can see that both classes exhibit some clumping but the structure of both classes is quite complex. Finally, in the fourth panel, we also include Fisher loss and entropy minimization which helps the two classes separate more in both domains.

\subsection{Model Interpretability: Visualizing Salient Regions in Input Images with Grad-CAMs}
Another way of probing deep neural network models is by identifying regions in the input images that proved most important for classification as a particular class. Domain adaptation should lead to differences in these important regions. In particular, without domain adaptation, the neural network can often identify incorrect or spurious regions in images it was not trained on in the target domain, while the classifier that works correctly should focus on regions that contain useful information for the given classification task.

Here we will use the Gradient-weighted Class Activation Mapping (Grad-CAM) method~\citep{SC2020} to visualize the regions which differently trained models identify as the most salient information in the image. This method calculates class $m$-specific gradients $\frac{\partial y_m}{\partial A_{ij}^l}$ of the output score $y_m$ with respect to the activation maps (i.e. feature maps) of the last convolutional layer $A_{ij}^l$ in the network. Here activation map dimensions are $i\times j = Z$ pixels and $l$ lists the number of feature maps. 
These gradients are global-average-pooled to calculate the importance weights $\alpha_m^l $ for a particular class $m$:
\begin{equation}
\alpha_m^l = \frac{1}{Z}\sum_{i}\sum_{j}\frac{\partial y_m}{\partial A_{ij}^l}.
\end{equation}
Grad-CAMs are then produced by applying a ReLU function (to extract positive activation regions for the particular class $m$) to the weighted combination of  $l$ feature maps in the last convolutional layer:
\begin{equation}
L_{\mathrm{Grad-CAM},m} = \mathrm{ReLU}\left(\sum_l \alpha_m^l A^l\right).
\end{equation}

\begin{figure*}
	\includegraphics[width=0.70\linewidth]{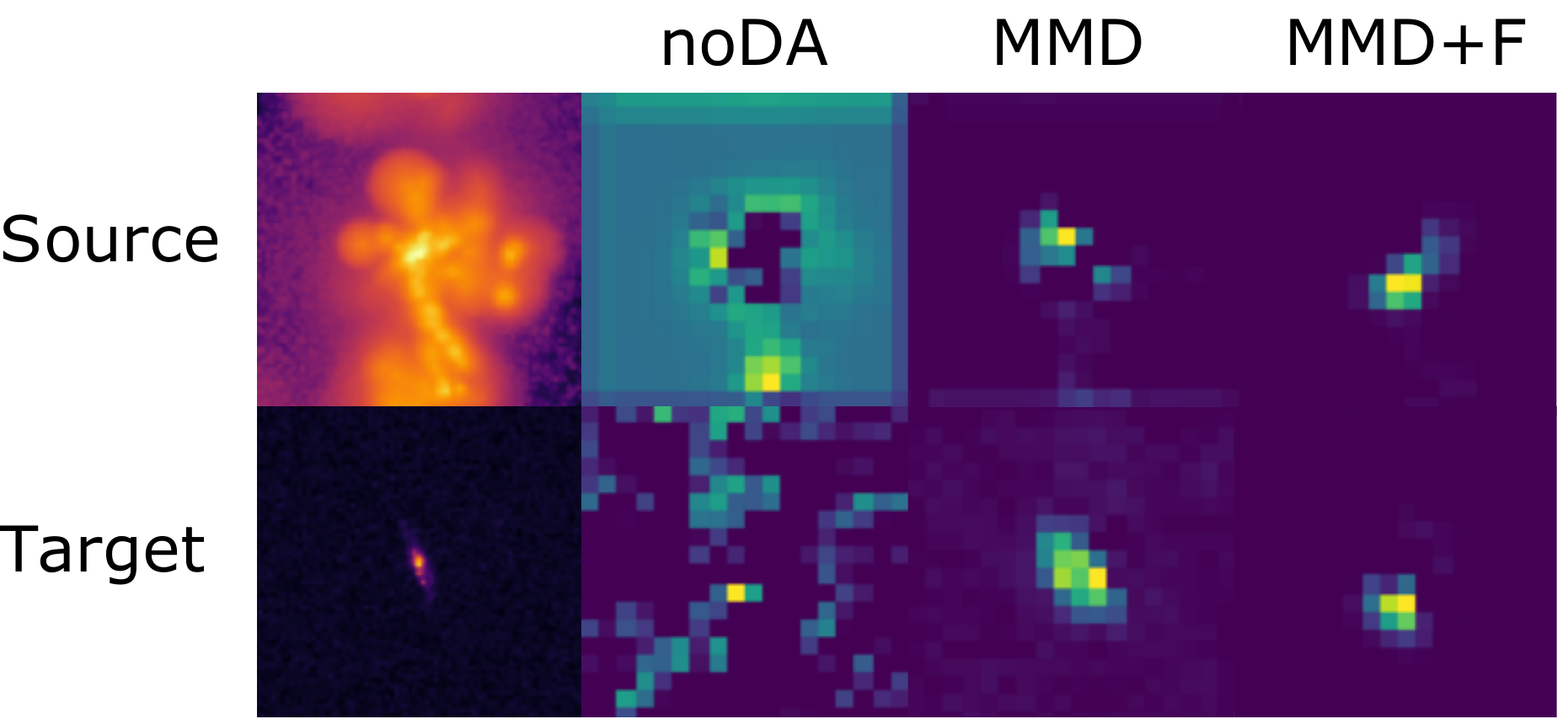}
    \caption{Grad-CAMs for simulation-to-simulation experiments in DeepMerge network made from feature maps in the final convolutional layer. The top left image shows an example galaxy merger image from the source domain (plotted with logarithmic colormap to enhance visibility), while the bottom left image is the same galaxy from the target domain. The second, third, and fourth column show Grad-CAMs for those images for classification into a merger class when training without domain adaptation, with MMD, and with MMD, Fisher loss, and entropy minimization, respectively. In the case of training without domain adaptation, the network is focusing on the periphery of the galaxy in the source domain network, and focuses on the noise in the target domain. When domain adaptation is introduced, the network learns to focus on the central brightest regions of the galaxy, which is visible in both domains.}
    \label{fig:gradcam}
\end{figure*}

In Figure~\ref{fig:gradcam}, we plot the last convolutional layer Grad-CAMs for simulation-to-simulation experiments with the DeepMerge network. We display plots for DeepMerge instead of ResNet18 due to the fact that the dimension of the last convolutional layer in ResNet18 is smaller, resulting in low-resolution Grad-CAMs that are much harder to interpret. The first column shows an example of a merging galaxy from the source domain at the top and from the target domain at the bottom; recall that these two domains differ only by the inclusion of noise. The second, third, and fourth columns show Grad-CAMs for the images in question for classification into a merger class, when training without DA, with MMD, and with MMD, Fisher loss, and entropy minimization, respectively.

In the case of training without domain adaptation in the second column, the network is focusing on the periphery of the galaxy in the source domain, exactly where a lot of interesting asymmetric and clumpy features are expected to appear in the case of mergers. These features are faint, and a lot of this information is lost in the target domain due to the inclusion of mimicked observational noise. As expected, the classifier does not work in the target domain: we can see that the network focuses on the noise instead. When domain adaptation is introduced--- MMD in the third row and MMD with Fisher loss and entropy minimization in the fourth column--- the network learns to focus on the central brightest regions of the galaxy, which are visible in both domains, and successfully performs classification in both cases.

\begin{figure*}
	\includegraphics[width=0.90\linewidth]{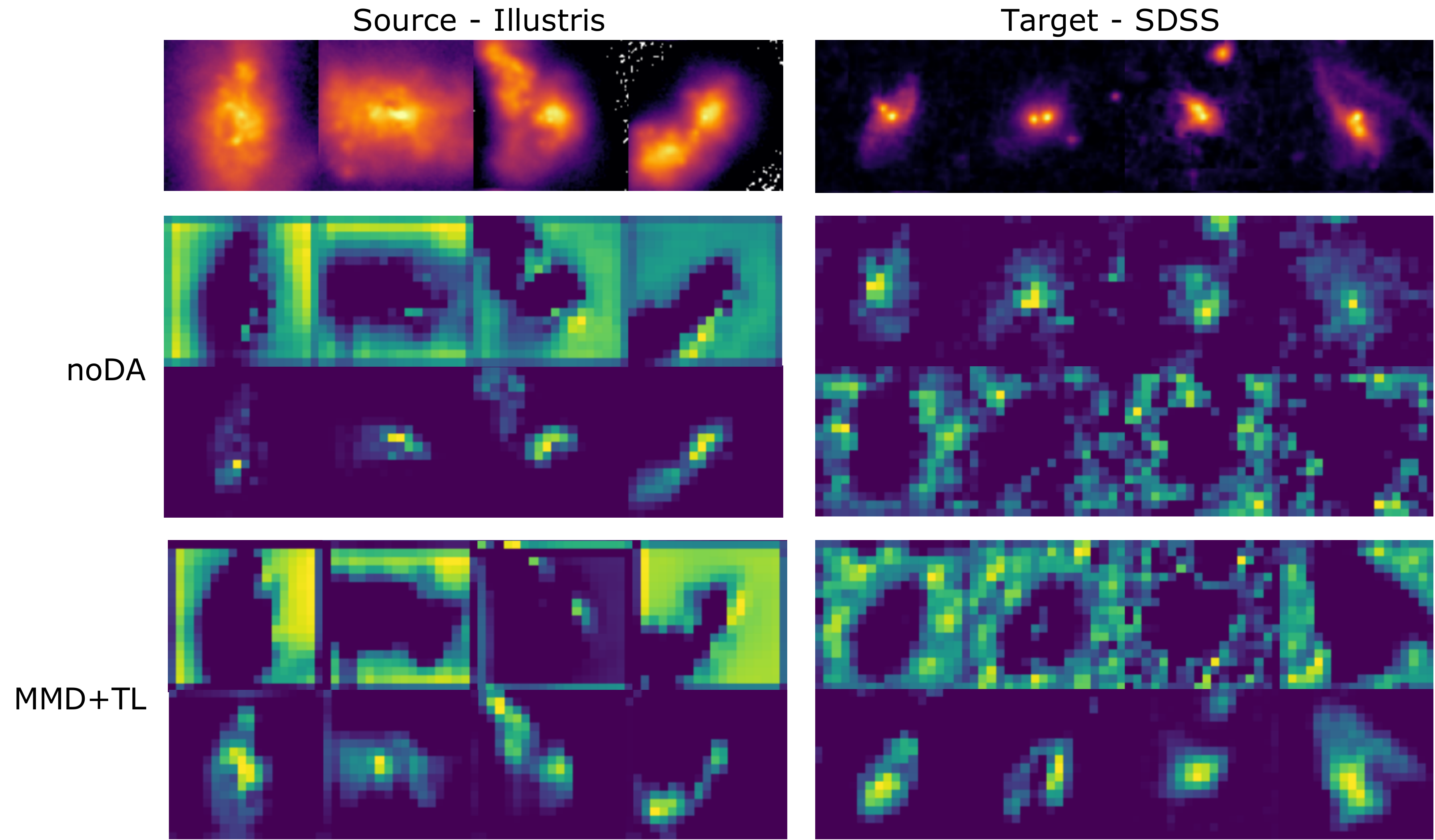}
    \caption{Grad-CAMs for simulation-to-real experiments in DeepMerge network. The top row of images shows examples of true mergers from simulated source dataset on the left and real target dataset on the right. The second row shows Grad-CAMs for these designated example images for classification into the merger class, while row three shows Grad-CAM for the same image for the non-merger class. We can see that, in case of the source domain, the peripheries are important for positive classification in the merger class, while central regions are important for positive classification as a non-merger. In the case of the target domain, both mergers and non-mergers look very different, so Grad-CAMs become noisy and display inverted behavior compared with the source domain. Finally, the third and fourth rows show merger and non-merger Grad-CAMs for the model trained with MMD and transfer learning. The successful domain adaptation is apparent, as the network performs both source and target domain classification in a similar manner as in the case of source domain classification without DA.}
    \label{fig:gradcam2}
\end{figure*}

Likewise, we plot Grad-CAMs for the simulated-to-real experiments with the DeepMerge network in Figure~\ref{fig:gradcam2}. Here we show multiple true merger images from the source and target domains in the top left and right columns, respectively. The second and third row show Grad-CAMs for these images when training without domain adaptation to highlight what the network focuses on for both merger and non-merger classes. Finally, the fourth and fifth rows show merger class and non-merger class Grad-CAMs for training with MMD with transfer learning. \\ 
\indent In the source domain Grad-CAMs--- where the classifier works with and without domain adaptation--- the neural network searches the periphery when classifying an example as a merger, while it focuses at the bright center when classifying non-mergers. This is to be expected, since mergers often have a lot of useful information on the periphery, while non-mergers are often very compact and only have a bright center in the middle of the image. On the other hand, the Grad-CAMs for the target domain without domain adaptation, i.e. for the unsuccessful classifier, demonstrate the network's focus on the noise. This even leads to the inverted characteristics from those described in the source domain: here classification as a merger depends on the bright center and classification as a non-merger depends on the peripheral information. This leads to the classifier failing to successfully distinguish mergers and non-mergers in the target domain. Finally, when training with MMD and transfer learning, the Grad-CAMs start to resemble those in the source domain and classification is successful in the target domain as well.

\subsection{Issues for Deployment in the Sciences: Avoiding Negative Transfer and Dealing With Small Datasets}
 
The ability to achieve success in combining cross-domain knowledge will largely depend on data availability and quality. We stress the importance of making efforts to achieve likeness between domains, particularly in the sciences. Most domain adaptation techniques were developed for use across very similar domains, such as pictures of office supplies in different settings in the \textit{Office-31} dataset~\citep{SK2010, HJ2017}. Scientists hoping to deploy these DA techniques face a much more nontrivial task, demonstrated by the challenges we encountered when trying to apply MMD and adversarial domain adaptation to the simulated-to-real datasets, which were both small and considerably different across domains.

Substantial domain divergence can result in suboptimal performance or even render domain adaptation techniques virtually useless. The presence of additional classes in the target domain or outliers within the same class make domain adaptation very difficult. Training a model on a demanding problem across dissimilar domains, which occurs frequently in the sciences, may lead to what is known as negative transfer. Negative transfer is when, rather than aiding domain adaptation, the knowledge learned from the source domain actually negatively impacts the performance of the classifier in the target domain. For a comprehensive survey of negative transfer and common methods to mitigate it see~\cite{ZD2020}. \\
\indent Therefore, despite its promise to help mitigate the issue of dealing with large unlabeled datasets, applying domain adaptation in the sciences may not be very straightforward. Scientists that wish to use DA techniques should strive to make their domains as similar as possible through careful dataset selection and image pre-processing (adding realistic noise, PSF, and other observational effects to simulated astronomical images). In crafting the simulated-to-real dataset in this paper, we made several choices to increase the likeness between our two domains, including choosing to use a small merger window to remove relaxed merger systems from our source domain, which were not present in our target domain. However, even in the case where our target distribution includes only post-mergers, DA was not very successful without the inclusion of transfer learning.\\
\indent Other clever approaches may be taken to decrease domain discrepancy. For example, it was shown in~\cite{BH2019} that standard deep learning algorithms trained to distinguish between galaxy merger stages in realistic data without domain adaption do not perform well when trained on pristine simulated images or even simulated images with included realistic PSF and noise. The authors achieve their best performance when simulated images were inserted into the realistic sky in order to introduce examples of crowding from nearby sources. This enabled the model to learn the distinction between crowding of nearby sources and close merging pairs. In this paper, the authors also show that achieving this observational realism is more important for good classification across domains than attempting to add more intricate details to computationally expensive simulations. \\
\indent Beyond dealing with negative transfer, data augmentation and transfer learning can aid substantially in training with small datasets. In this paper, we demonstrated that a combination of MMD and transfer learning from our simulated-to-simulated dataset enhanced the learning of correct features in our extremely small simulated-to-real dataset. Even in the case of small and quite discrepant domains, domain adaptation techniques can be successfully used to improve performance of deep learning algorithms on unlabeled observational data.\\
\indent In the future, more refined domain adaptation techniques will likely be needed in the sciences. In the all too prevalent case when classes look very different across source and target domains, future work will include applying methods such as class-aware Contrastive Domain Discrepancy networks~\citep{KJ2019}, that promote class compactness and overlap between class distributions from different domains--- rather than the overlap of entire source and target distributions performed with MMD--- to achieve even higher accuracies in the target domain.\\
\indent Despite the issues mentioned above, we maintain that the prospect for domain adaptation's use in the sciences is extremely promising. With ongoing domain adaptation development both within computer science and by those who leverage it creatively in other sciences, we are confident that DA will soon become a staple in the natural scientist's toolkit to leverage all available data.

\section{Conclusion}
\label{sec:conclusion} 

In this paper, we focus on applying domain adaptation techniques to the astronomical context of studying galaxy mergers. Galaxy mergers are crucial in the study of galaxy morphology, evolution, star formation, as well as particle acceleration and the evolution of matter in the universe. Finding comprehensive samples of merging galaxies in different merger stages is very important for the study of these long processes, and we are excited about what the next era of large-scale survey data will bring.

MMD and adversarial training using DANNs show great promise for use in classifying galaxy mergers across domains. We showed here how they can help improve classification accuracies in an unlabeled target domain, thereby allowing models trained on simulated labeled data to be successfully applied on both mimicked and real observational data. While we demonstrate successful implementation of both methods in the simulated-to-simulated dataset for both DeepMerge and ResNet18, we also present promising results for MMD combined with transfer learning in the simulated-to-real dataset in DeepMerge. In both types of experiments we were able to increase the target dataset accuracy in the testing phase by up to $\sim20\%$.

While we found that MMD and adversarial training can be challenging to fine-tune for nontrivial scientific tasks, we conclude that domain adaptation techniques will soon flourish as a necessary tool in astronomy and other natural sciences. These techniques will play an important role in successful use of deep learning algorithms on huge datasets from future large astronomical surveys, and might even help in real-time detection of transient objects and other interesting phenomena. We affirm that domain adaptation techniques will prove essential to building deep learning models that can combine and harness all available observational and simulated data, a tantalizing prospect in the sciences.

\section*{Acknowledgements}
This manuscript has been supported by Fermi Research Alliance, LLC under Contract No. DE-AC02-07CH11359 with the U.S. Department of Energy, Office of Science, Office of High Energy Physics. This research has been partially supported by the High Velocity Artificial Intelligence grant as part of the Department of Energy High Energy Physics Computational HEP sessions program. This research used resources of the Argonne Leadership Computing Facility at Argonne National Laboratory, which is a user facility supported by the Office of Science of the U.S. Department of Energy under contract DE-AC02-06CH11357.

The authors of this paper have committed themselves to performing this work in an equitable, inclusive, and just environment, and we hold ourselves accountable, believing that the best science is contingent on a good research environment.
We acknowledge the Deep Skies Lab as a community of multi-domain experts and collaborators who have facilitated an environment of open discussion, idea-generation, and collaboration. This community was important for the development of this project.

We also thank K. Pedro, N. Tran, W.~J. Pearson, Y. Zhang and M. Vasist for valuable discussion and comments.

\subsection*{\it{Author Contributions}}

A.~\'Ciprijanovi\'c--- \textit{Conceptualization, Data curation, Formal analysis, Investigation, Methodology, Resources, Software, Visualization, Project administration, Supervision, Writing of original draft}; D.~Kafkes--- \textit{Formal analysis, Investigation, Methodology, Resources, Software, Visualization, Writing of original draft}; K.~Downey--- \textit{Data curation, Formal analysis, Investigation, Software, Visualization}; S.~Jenkins--- \textit{Formal analysis, Investigation, Software, Visualization, Writing (review \& editing)}; G.~N.~Perdue--- \textit{Investigation, Methodology, Project administration, Resources, Software, Supervision, Writing  (review \& editing)}; S.~Madireddy--- \textit{Resources, Software, Methodology, Supervision, Writing (review \& editing)}; T.~Johnson--- \textit{Methodology, Supervision, Writing (review \& editing)}; G.~F.~Snyder---\textit{Methodology, Conceptualization, Data curation, Writing (review \& editing)}; B.~Nord--- \textit{Methodology, Conceptualization, Supervision, Writing (review \& editing)}.

\section*{Data and Code Availability}

All simulated Illustris datasets and observed SDSS dataset are available on \href{https://doi.org/10.5281/zenodo.4507941}{Zenodo}. For code see our \href{https://github.com/AleksCipri/DeepMergeDomainAdaptation}{GitHub} page.


\bibliographystyle{mnras}
\bibliography{main} 

\clearpage


\appendix

\section{Neural Network Hyperparameters and Training Details}
\label{sec:networks_params}

Here we list all relevant details related to training the neural networks used in this paper. In Table~\ref{table:arch}, we give details about the architecture of DeepMerge. For ResNet18 architecture, see ~\cite{HZ2015}.\\
\indent For both DeepMerge and ResNet18, we performed hyperparameter optimization experiments with DeepHyper~\citep{BS2018,BE2019}, an open source mixed-integer nonlinear optimization framework which employs a parallel asynchronous-model-based search approach to find the high-performing parameter configuration in this mixed (categorical, continuous, integer) search space. Our objective is to minimize the total loss ${\cal L}_\mathrm{TOT}$ for MMD hyperparameter searches and to minimize a more complex objective for adversarial domain adaptation: ${\cal L}_\mathrm{CL}(1 + \Delta{\cal A}_\mathrm{s} + \Delta{\cal A}_\mathrm{t})$, where $\Delta {\cal A}_\mathrm{s} = (0.5 - {\cal A}_\mathrm{s})^2$  and $\Delta {\cal A}_\mathrm{t} = (0.5 - {\cal A}_\mathrm{t})^2$. This was done in an attempt to prevent domain classifier mode collapse (all images classified as one domain) and limit how much the domain classifier's source domain accuracy ${\cal A}_\mathrm{s}$ and target domain accuracy ${\cal A}_\mathrm{t}$ differed from $0.5$, the ideal result for a perfectly confused domain classifier.\\
\indent Different hyperparameters were needed for different experiments, i.e. their values are highly dependent not only on the network, but also on the domain adaptation technique used and the dataset in question. We list all hyperparameters used in the simulated-to-simulated experiments in Table~\ref{table:params} and the simulated-to-real experiments in Table~\ref{table:params2}.\\
\indent For training both DeepMerge and ResNet18, we use the Adam optimizer~\citep{KB14}. Additionally, we implement "one-cycle" learning rate scheduling~\citep{ST2018}, which splits a given cycle length in half and includes equal length linear scaling up and down of values from a minimum to a maximum value (like a sawtooth pattern). This technique offers the best of both worlds: higher learning rates assist in regularization by enabling egress from saddle points and lower learning rates prevent the training from diverging ~\cite{ST2018}. It was shown by~\cite{ST2018} that this one-cycle learning leads to much faster convergence of training accuracy. If present within the chosen optimizer, momentum also follows this scaling with optional annihilation. If specified, annihilation requires rapid scaling to small learning rate values following the cycling down period in order to enable a steeper fall into a local minimum in the loss-landscape. \\
\indent Following ~\cite{ST2018}, the choice of maximum and minimum learning rates for this technique were performed using a Learning Rate Range Scan in which we linearly increased values over several orders of magnitude during the first few epochs of training and plotted the associated total loss against the epoch. The minimum of these curves was taken as the maximum value of the learning rate, with the minimum learning rate taken as one order of magnitude lower than the maximum. As for cycle length, results of~\cite{ST2018} suggest that learning should be relatively robust to choices of between 2-10 epochs per cycle length. Here, we select cycle length through the aforementioned DeepHyper hyperparameter search for both DeepMerge and ResNet18.\\
\indent In the case of domain adversarial training, we found the performance of the domain classifier to be quite finicky: mode collapse in the domain classifier was common despite training achieving high classification accuracy in the base network. We added the ability to multiply the learning rate of this part of the network by a user-defined constant in order to implement differential learning rates across the networks. For tasks similar to ours, we believe that this factor should often be less than one--- indeed small fractions proved quite helpful to the domain adversarial training, especially with Fisher loss included.

\begin{table*}
  \centering
  \noindent\begin{minipage}[b]{0.99\textwidth}
   \centering
    \caption{Architecture of the DeepMerge CNN used in this paper.}
  \label{table:arch}
  \centering
  \begin{tabular}{|l | l l l  l   l |}
\hline
\bf{Layers}         & \bf{Properties}           & \bf{Stride}       & \bf{Padding}  & \bf{Output Shape} & \bf{Parameters}   \\ \hline\hline
Input               & $3\times75\times75$\footnote{We use "channel first" image data format.}       & -                 & -             & (3, 75, 75)       & 0                 \\ \Cline{1.2pt}{1-6}
Convolution (2D)    & Filters: 8                & $1$        & $2$          & (8, 75, 75)       & 608               \\
                    & Kernel: $ 5\times5$       & -                 & -             & -                 & -                 \\
                    & Activation: ReLU          & -                 & -             & -                 & -                 \\ \hline
Batch Normalization & -                         & -                 & -             & (8, 75, 75)       & 16               \\ \hline
MaxPooling          & Kernel: $2\times2$        & $2$        & 0         & (8, 37, 37)       & 0                 \\ \hline
Convolution (2D)    & Filters: 16               & $1$        & $1$          & (16, 37, 37)      & 1168              \\
                    & Kernel: $ 3\times3$       & -                 & -             & -                 & -                 \\
                    & Activation: ReLU          & -                 & -             & -                 & -                 \\ \hline
Batch Normalization & -                         & -                 & -             & (16, 37, 37)      & 32               \\ \hline
MaxPooling          & Kernel: $2\times2$        & $2$        & 0         & (16, 18, 18)      & 0                 \\ \hline
Convolution (2D)    & Filters: 32               & $1$        & 1          & (32, 18, 18)      & 4640              \\
                    & Kernel: $ 3\times3$       & -                 & -             & -                 & -                 \\
                    & Activation: ReLU          & -                 & -             & -                 & -                 \\ \hline
Batch Normalization & -                         & -                 & -             & (32, 18, 18)      & 64                \\ \hline
MaxPooling          & Kernel: $2\times2$        & $2$        & 0         & (32, 9, 9)        & 0                 \\ \hline
Flatten             & -                         & -                 & -             & (2592)            & -                 \\ \hline
Fully connected     &   Activation: ReLU          & -                 & -             & (64)              & 165952            \\ \hline
Fully connected     &      Activation: ReLU     & -                 & -             & (32)              & 2080              \\ \hline
Fully connected     & Activation: Softmax       & -                 & -             & (2)               & 66                \\ \hline
\end{tabular}
\end{minipage}
\end{table*}

\begin{table*}
   \centering
   \noindent\begin{minipage}[b]{0.99\textwidth}
   \centering
    \caption{The hyperparameters used to train in the simulated-to-simulated dataset experiments. The third and fourth columns list parameters for DeepMerge and ResNet18, respectively. The first row shows parameters for our baseline case without domain adaptation, while all other rows give parameters used in different domain adaptation experiments.}
  \label{table:params}
  \centering
  \begin{tabular}{|l | l |c|c|}
    \multicolumn{1}{c}{}  & \multicolumn{3}{c}{Simulated-to-Simulated}  \\\hline 
Experiment      &   Hyperparameters  &  DeepMerge   &   ResNet18 \\\Cline{1.2pt}{1-4}
\multirow{5}{*}{No Domain Adaptations}    &  Learning rate              &  $0.001$      & $0.00023$            \\
                                          &  Beta                       &  $(0.7, 0.8)$ & $(0.7, 0.8)$             \\
                                          &  Weight Decay               &  $0.01$       & $0.1$                           \\
                                          &  Epsilon                    &  $10^{-8}$    & $10^{-8}$                   \\
                                          &  Cycle Length               &  $8$          & $2$                             \\
                                          &  Early stopping patience    &  $20$         & $15$                        \\
                                          &  Dropout                    &  N/A          & $0.6 $                         \\ \hline
\multirow{5}{*}{MMD}                      &  Learning rate              &  $0.001$      & $0.00023$                            \\
                                          &  Beta                       &  $(0.7, 0.8)$ & $(0.7, 0.8)$                      \\
                                          &  Weight Decay               &  $0.0001$     & $0.08$                              \\
                                          &  Epsilon                    &  $10^{-8}$    & $10^{-8}$                       \\
                                          &  Cycle Length               &  $8$          & $2$                               \\
                                          &  Early stopping patience    &  $20$         & $15$                               \\
                                          &  $\lambda_\mathrm{TL}$      &  $1$          & $0.01$                              \\\hline
\multirow{5}{*}{MMD + Fisher + Entropy}   &  Learning rate              &  $0.001$      & $0.00023$                      \\
                                          &  Beta                       &  $(0.7, 0.8)$ & $(0.7, 0.8)$                     \\ 
                                          &  Weight Decay               &  $0.0001$     & $0.08$                          \\
                                          &  Epsilon                    &  $10^{-8}$    & $10^{-8}$                        \\
                                          &  Cycle Length               &  $8$          & $2$                             \\
                                          &  Early stopping patience    &  $20$         & $15$                               \\
                                          &  $\lambda_\mathrm{TL}$      &  $1$          & $1$                            \\
                                          &  $\lambda_\mathrm{w}$       &  $0.01$       & $.01$                     \\
                                          &  $\lambda_\mathrm{b}$       &  $1.0$        & $1.0$                          \\
                                          &  $\lambda_\mathrm{EM}$      &  $0.05$       & $0.05$                        \\\hline
\multirow{5}{*}{Adversarial}              &  Learning rate              &  $0.001$      & $0.001$                         \\
                                          &  Beta                       &  $(0.7, 0.8)$ & $(0.7, 0.8)$                   \\ 
                                          &  Weight Decay               &  $0.0001 $    & $0.0001$                    \\
                                          &  Epsilon                    &  $10^{-8}$    & $10^{-8}$                        \\
                                          &  Cycle Length               &  $8$          & $2$                              \\
                                          &  Early stopping patience    &  $20$         & $15$                              \\
                                          &  $\lambda_\mathrm{TL}$      &  $1$          & $0.1$                             \\
                                          &  Domain class. LR mult.     &  $1$          & $1$                       \\\hline
\multirow{5}{*}{Adversarial + Fisher + Entropy}    & Learning rate      &  $0.001$      & $0.001$                       \\
                                          &  Beta                       &  $(0.7, 0.8)$ & $(0.7, 0.8)$                \\ 
                                          &  Weight Decay               &  $0.0001 $    & $0.0001$                          \\
                                          &  Epsilon                    &  $10^{-8}$    & $10^{-8}$                          \\
                                          &  Cycle Length               &  $8$          & $8$                                  \\  
                                          &  Early stopping patience    &  $20$         & $15$                               \\
                                          &  $\lambda_\mathrm{TL}$      &  $1$          & $0.1$                             \\
                                          &  $\lambda_\mathrm{w}$       &  $0.0001$     & $0.01$                               \\
                                          &  $\lambda_\mathrm{b}$       &  $1$          & $1$                                \\
                                          &  $\lambda_\mathrm{EM}$      &  $0.0001$     & $0.05$                               \\
                                          & Domain class. LR mult.      &  $0.01$       & $0.1$                          \\\hline
\end{tabular}
\end{minipage}
\end{table*}

\begin{table*}
   \centering
   \noindent\begin{minipage}[b]{0.70\textwidth}
   \centering
    \caption{The hyperparameters used to train DeepMerge on the simulated-to-real dataset. The second column of the table shows values for the baseline without domain adaptation, while the third column gives parameters for MMD and the fourth for MMD with transfer learning from the simulated-to-simulated dataset.}
  \label{table:params2}
  \centering
  \begin{tabular}{| l | c | c | c |}
    \multicolumn{1}{c}{}   & \multicolumn{3}{c}{DeepMerge: Simulated-to-Real}\\\hline 
Hyperparameters & noDA  & MMD & TL+MMD     \\\Cline{1.2pt}{1-4}
Learning rate        &   $0.001$    &   $0.001$    &  $0.01$   \\
Beta                 &   $(0.7, 0.8)$    &  $(0.7, 0.8)$     &    $(0.7, 0.8)$           \\
Weight Decay         &  $0.001$     &  $0.001$     &    $0.0001$         \\
Epsilon              & $10^{-8}$      &   $10^{-8}$    &  $10^{-8}$            \\
Cycle Length         &   $5$     &    $5$   &     $5$         \\
Early stopping patience    &  $20$  &   $20$     &    $20$       \\\hline
\end{tabular}
\end{minipage}
\end{table*}

All experiments were run on Google Colab GPU instances and Google Console virtual machine with a NVIDIA Tesla T4 GPU. Our code uses the PyTorch framework and a fixed random seed=1; it is important to note that completely reproducible results are not guaranteed across PyTorch releases or between CPU and GPU executions, even when using fixed random seeds. In the case of training with GPUs, some PyTorch functions that use CUDA can introduce non-deterministic results. We make sure that this is fixed in the backend so that rerunning the code always leads to the same results. Yet we still must note that, even though our code will produce deterministic results when running on a single machine, slight differences will be present in the case of running on different machines.

\section{More on Network Performance}
\label{sec:networks_results}

In this section we provide more details to further compare both DeepMerge and Resnet18 performance in experiments with and without domain adaptation. To complement the results presented in Section~\ref{sec:results_ss} for simulated-to-simulated experiments, in Table~\ref{table:CMs} we compactly display confusion matrices--- for the test phase using both source and target data--- for all simulated-to-simulated experiments performed with DeepMerge at the top and ResNet18 at the bottom.

\begin{figure*}
    \includegraphics[width=0.99\linewidth]{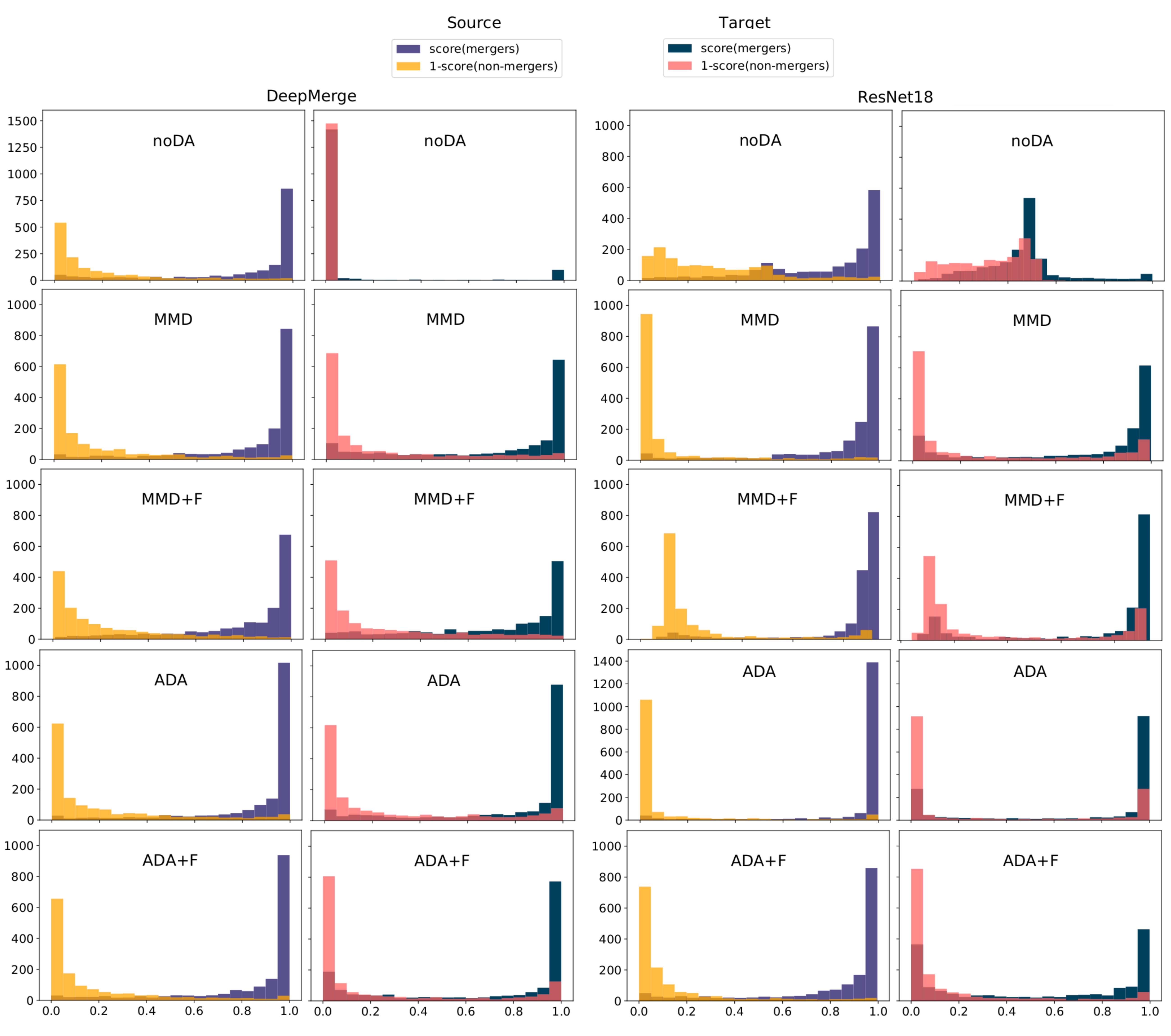}
    \caption{Simulated-to-Simulated: Histograms of the output scores of DeepMerge (left two columns) and ResNet18 (right two columns) for the test set of images with the source domain (simulated pristine images) on the left--- mergers as dark purple and non-mergers as yellow--- and target domain (simulated noisy images) on the right--- mergers as navy blue and non-mergers as pink. From top to bottom, results are given for training without domain adaptation, MMD, MMD with Fisher loss and entropy minimization, adversarial training, and adversarial training with Fisher loss and entropy minimization. We plot the histogram of the output scores for all images that represent true mergers, and 1-score for all non-merger images in order to separate the classes for better visibility. Note that the vertical axis range is the same for all experiments except for no domain adaptation for DeepMerge and adversarial domain adaptation for ResNet18 (in order to accommodate larger bars).}\label{fig:histograms-SimSim}
\end{figure*}

We also plot histograms of the output scores for DeepMerge in Figure~\ref{fig:histograms-SimSim} (left two columns) and ResNet18 (right two columns) for the simulated-to-simulated dataset. In these figures, the source domain is on the left with mergers as dark purple and non-mergers as yellow, and the target domain is on the right with mergers as navy blue and non-mergers as pink. Histograms from top to bottom are ordered as: no domain adaptation, MMD, MMD with Fisher loss and entropy minimization, adversarial domain adaptation, and adversarial domain adaptation with Fisher loss and entropy minimization.

From Figure~\ref{fig:histograms-SimSim}, it is clear that adversarial training, for both DeepMerge and ResNet18, leads to outputs more tightly concentrated around $0$ and $1$, i.e. the network seems more confident about the classification, while using MMD leads to slightly more spread out results. Still, confidence is not a guarantee of best classification accuracy and indeed ResNet18 performs the best in the target domain with MMD with Fisher loss and entropy minimization (although, we note that overall the differences in performance between different methods are not large). Also, in the case of the labeled source domain, it is evident that the inclusion of additional losses works as a regularization mechanism, allowing the network to train for longer without overfitting. This effect is particularly noticeable in the case of ResNet18, which is a larger network and thus more prone to overfitting. As a result, in the source domain, we can see that the ResNet18 histograms have more outputs very close to $0$ or $1$ in experiments with domain adaptation, than without it.

Using the confusion matrices and histograms for the simulated-to-simulated experiments we notice that, depending on the network, inclusion of different losses leads to different behaviours. In the case of DeepMerge, the best performing model uses adversarial training. The inclusion of Fisher loss and entropy minimization with this technique leads to slightly worse performance. On the other hand, in the case of ResNet18, the best performance is the result of using MMD with Fisher loss and entropy minimization. When these additional losses are added to the adversarial trained ResNet18, performance drops significantly: the merger class gets classified incorrectly $46\%$ of the time.

More details related to the results presented in Section~\ref{sec:results_sr} for simulated-to-real experiments can be found in Table~\ref{table:CMs_rs}, where we present confusion matrices for classification of test source and target datasets with DeepMerge in the case of training without domain adaptation, with MMD, and with MMD with transfer learning.

\begin{figure*}
    \includegraphics[width=.65\linewidth]{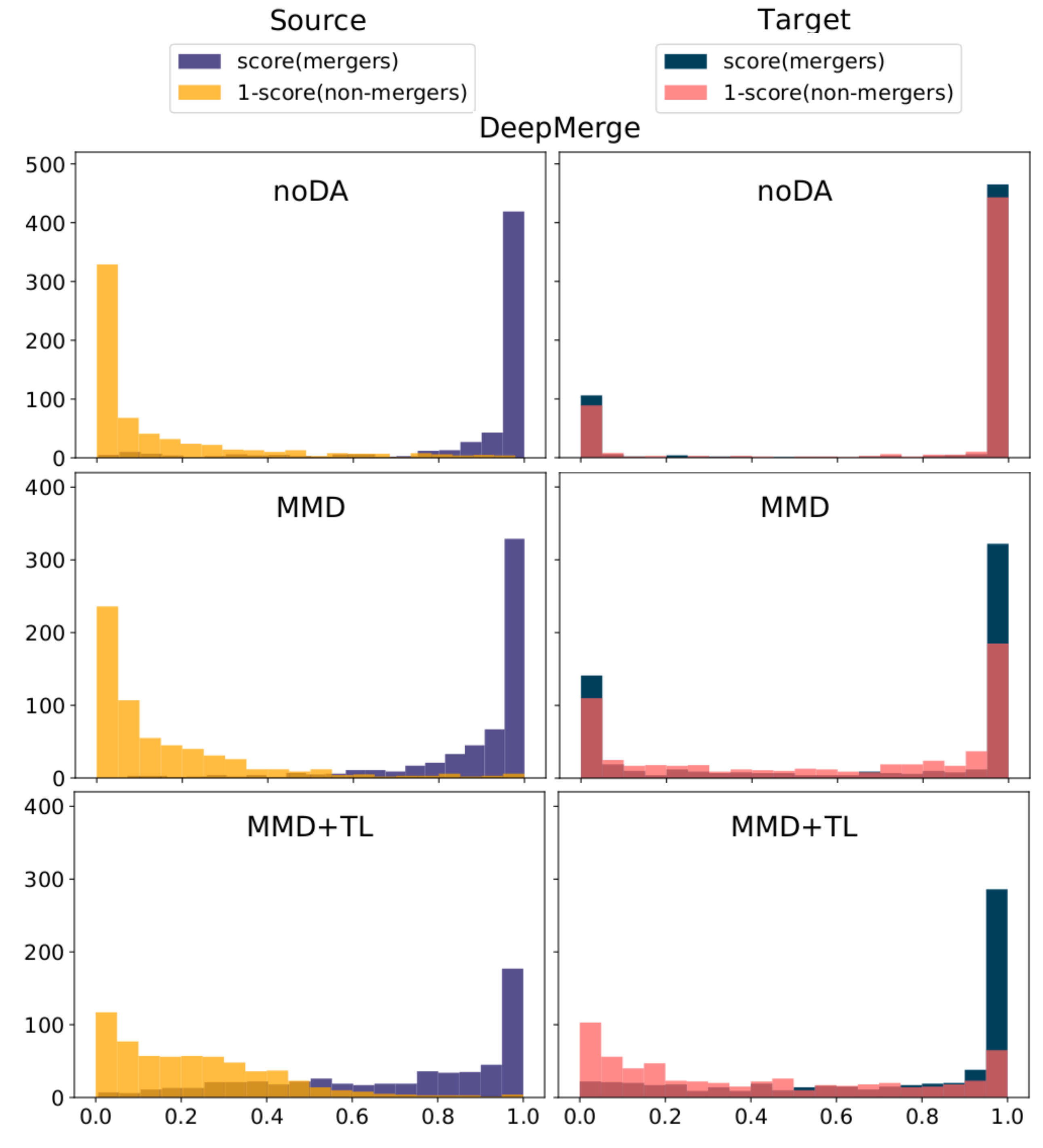}
    \caption{Simulated-to-Real: Histograms of the output scores of DeepMerge for the test set of images with the source domain on the left (simulated Illustris images)--- mergers as dark purple and non-mergers as yellow--- and target domain (real SDSS images) on the right--- mergers as navy blue and non-mergers as pink. From top to bottom, results are given for training without domain adaptation, MMD, and MMD with transfer learning. We plot histograms of the output scores for all images that represent true mergers, and 1-score of all non-merger images in order to separate the classes for better visibility. Note that the vertical axis range is the same for all experiments except for no domain adaptation (in order to accommodate larger bars).}\label{fig:histograms-SDSS}
\end{figure*}

We also plot histograms of the output scores for simulated-to-real experiments in Figure~\ref{fig:histograms-SDSS}--- from top to bottom we present no domain adaptation, MMD, and MMD with transfer learning, respectively. It is clear that the addition of MMD results in a broadening of the source histogram distributions, especially for the non-merger class. When transfer learning is employed, distributions are further spread. This provides a visual of how a network trained with domain adaption is restricted to learn the set of domain-invariant features, rather than exploiting all available features in the source domain, which would result in greater confidence. Meanwhile, in the target domain, we see that classifier does not work without DA and with simple MMD, with many examples from both classes classified incorrectly. Inclusion of transfer learning allows correct classification even in the target domain, but with large spread of the output values, especially for the non-merger class. 

Finally, we checked the stability of neural network performance to the choice of the random seed. To test this, we trained the DeepMerge architecture, trained on each dataset pair, ten times using ten different seeds for each type of experiment. Table~\ref{table:seeds} shows the means and standard deviations for all relevant reported performance metrics for the simulation-to-simulation dataset of distant merging galaxies. Since our dataset is very small in the simulated-to-real experiments, we expect a larger spread in performance metric results in the testing stage when different random seeds are used. In Table~\ref{table:seeds_rs}, we again give the mean and standard deviations of different performance metrics, in order to show this slightly larger spread.

\begin{table*}
   \centering
   \noindent\begin{minipage}[b]{0.99\textwidth}
   \centering
    \caption{Normalized confusion matrices for simulated-to-simulated experiments. The top table gives results for DeepMerge, while ResNet18 is presented in the bottom table. Here the true labels are presented horizontally and the predicted labels are given vertically. Finally, in each table, the top row shows confusion matrices for the source domain test set of images, while the bottom row shows results from the target test dataset.}
  \label{table:CMs}
  \centering
  \begin{tabular}{ c |c | c | c c | c c | c c | c c | c c | c c | c c | c c | c c | c c |}
  \multicolumn{3}{c}{}  & \multicolumn{10}{c}{DeepMerge: Simulated-to-Simulated}\\\cline{2-13}
\multirow{2}{*}{} & \multicolumn{2}{c|}{Experiment}  & \multicolumn{2}{c|}{noDA}  &  \multicolumn{2}{c|}{MMD} &  \multicolumn{2}{c|}{MMD + F} &  \multicolumn{2}{c|}{ADA}  &  \multicolumn{2}{c|}{ADA + F} \\\cline{2-13}
  & \multicolumn{2}{c|}{True Label}  & M  &  NM & M &  NM &  M  &  NM &  M  &  NM &  M  &  NM  \\\Cline{1.2pt}{2-13}
\multirow{2}{*}{Source} & \multirow{4}{*}{Predicted Label} &  M  & $0.83$      &  $0.13$   &   $0.87$  & $0.14$   &  $0.84$  &    $0.15$   &  $0.89$    & $0.15$       &  $0.86$     &    $0.12$   \\
  &  &  NM  &  $0.17$     &  $0.87$   &  $0.13$   & $0.86$   &  $0.16$  &  $0.85$      &    $0.11$  & $0.85$       &    $0.14$   &  $0.88$      \\\cline{3-13}
\multirow{2}{*}{Target} &  &  M  &   $0.08$    &  $0.0$   & $0.72$    & $0.19$   &  $0.75$  &    $0.21$   &   $0.81$   &     $0.24$   &   $0.69$    &    $0.21$     \\
  & &  NM     &   $0.92$    & $1.0$    &  $0.28$   & $0.81$   &  $0.25$  & $0.79$      &  $0.19$    & $0.76$       &   $0.31$    & $0.79$         \\\cline{2-13}
\multicolumn{13}{c}{}  \\
\multicolumn{13}{c}{}  \\
\multicolumn{13}{c}{}  \\
\multicolumn{3}{c}{}  & \multicolumn{10}{c}{ResNet18: Simulated-to-Simulated}\\\cline{2-13}
\multirow{2}{*}{} & \multicolumn{2}{c|}{Experiment}  & \multicolumn{2}{c|}{noDA}  &  \multicolumn{2}{c|}{MMD} &  \multicolumn{2}{c|}{MMD + F} &  \multicolumn{2}{c|}{ADA}  &  \multicolumn{2}{c|}{ADA + F} \\\cline{2-13}
  & \multicolumn{2}{c|}{True Label}  & M  &  NM & M &  NM &  M  &  NM &  M  &  NM &  M  &  NM  \\\Cline{1.2pt}{2-13}
\multirow{2}{*}{Source} & \multirow{4}{*}{Predicted Label} &  M  &  $0.83$     & $0.22$    &  $0.90$   & $0.09$   &  $0.90$  & $0.12$     &  $0.93$    &  $0.08$      &  $0.85$     &   $0.09$   \\
  &  &  NM  & $0.17$   & $0.78$   &  $0.10$  & $0.91$   &  $0.10$  &  $0.88$     & $0.07$     &  $0.92$      &  $0.15$     &   $0.91$     \\\cline{3-13}
\multirow{2}{*}{Target} &  &  M  & $0.31$     & $0.07$     & $0.74$   & $0.27$   & $0.78$  & $0.28$      & $0.71$    & $0.28$      &  $0.54$    & $0.12$        \\
  & &  NM     &   $0.69$   & $0.93$   &  $0.26$   & $0.73$   & $0.22$   &  $0.72$    &  $0.29$    &$0.72$ & $0.46$      &  $0.88$      \\\cline{2-13}

\end{tabular}
\end{minipage}
\end{table*}

\begin{table*}
   \centering
   \noindent\begin{minipage}[b]{0.99\textwidth}
   \centering
    \caption{Results from running DeepMerge experiments with ten different random seeds. Seeds are used for image shuffling, weight initialization, and CUDA backend. We present means and standard deviations for all aforementioned performance metrics. We do not see substantial variation within each experiment.}
  \label{table:seeds}
  \centering
  \begin{tabular}{|l | l | c c| }
    \multicolumn{1}{c}{}  & \multicolumn{3}{c}{DeepMerge: Simulated-to-Simulated}\\\hline 
Experiment    &  Metric     &  Source   &   Target  \\\Cline{1.2pt}{1-4}
\multirow{5}{*}{No Domain Adaptation}           &  AUC          &   $0.91\pm0.004$        &  $0.72\pm0.03$   \\
                                                &  Accuracy     &   $0.84\pm0.01$         & $0.57\pm0.05$     \\ 
                                                &  Precision    &   $0.86\pm0.05$         &   $0.84\pm0.17$   \\
                                                &  Recall       &   $0.83\pm0.03$         & $0.38\pm0.33$  \\
                                                &  F1 score     &   $0.84\pm0.01$         &  $0.40\pm0.24$     \\
                                                &  Brier score  &   $0.12\pm0.004$         &  $0.40\pm0.07$   \\\hline
\multirow{5}{*}{MMD}                            &  AUC          &   $0.93\pm0.01$        &  $0.86\pm0.01$     \\
                                                &  Accuracy     & $0.86\pm0.01$          & $0.77\pm0.01$     \\ 
                                                &  Precision    &   $0.87\pm0.02$        &  $0.81\pm0.02$    \\
                                                &  Recall       &     $0.88\pm0.01$      &  $0.75\pm0.03$     \\
                                                &  F1 score     &    $0.87\pm0.01$       &  $0.78\pm0.01$     \\
                                                &  Brier score  &  $0.10\pm0.01$         &  $0.16\pm0.005$      \\\hline
\multirow{5}{*}{MMD + Fisher + Entropy}         &  AUC          &    $0.92\pm0.01$        &  $0.85\pm0.01$     \\
                                                &  Accuracy     &  $0.84\pm0.01$         & $0.77\pm0.01$        \\ 
                                                &  Precision    &   $0.84\pm0.02$        &  $0.80\pm0.02$     \\
                                                &  Recall       &   $0.84\pm0.03$        &  $0.75\pm0.05$       \\
                                                &  F1 score     &    $0.85\pm0.01$       &  $0.77\pm0.02$     \\
                                                &  Brier score  &   $0.12\pm0.01$        &  $0.16\pm0.01$      \\\hline
\multirow{5}{*}{Adversarial}                    &  AUC          &   $0.94\pm0.01$         &  $0.87\pm0.01$    \\
                                                &  Accuracy     & $0.86\pm0.01$           & $0.79\pm0.005$       \\ 
                                                &  Precision    &    $0.87\pm0.02$       &  $0.80\pm0.02$      \\
                                                &  Recall       &    $0.88\pm0.02$       &  $0.80\pm0.03$     \\
                                                &  F1 score     &    $0.87\pm0.01$       &   $0.80\pm0.01$     \\
                                                &  Brier score  &  $0.10\pm0.006$         &  $0.15\pm0.005$     \\\hline
\multirow{5}{*}{Adv. + Fisher + Entropy}        &  AUC          &     $0.94\pm0.01$      &  $0.84\pm0.01$    \\
                                                &  Accuracy     &     $0.87\pm0.01$       & $0.75\pm0.01$        \\ 
                                                &  Precision    &   $0.88\pm0.02$         &  $0.76\pm0.05$         \\
                                                &  Recall       &    $0.88\pm0.02$       &  $0.78\pm0.07$      \\
                                                &  F1 score     &  $0.88\pm0.01$         &  $0.77\pm0.02$      \\
                                                &  Brier score  &  $0.10\pm0.01$         &  $0.19\pm0.02$    \\\hline
\end{tabular}
\end{minipage}
\end{table*}

\begin{table*}
   \centering
   \noindent\begin{minipage}[b]{0.99\textwidth}
   \centering
    \caption{Normalized confusion matrices for simulated-to-real experiments with DeepMerge. True labels are presented horizontally, while predicted labels are vertical. Finally, the top row shows confusion matrices for the source domain test set of images, while the bottom row give results of the classification of the target test dataset.}
  \label{table:CMs_rs}
  \centering
  \begin{tabular}{ c |c | c | c c | c c | c c | c c | c c | c c | c c | c c |}
\multicolumn{3}{c}{}  & \multicolumn{8}{c}{DeepMerge: Simulated-to-Real}\\\cline{2-9}
\multirow{2}{*}{} & \multicolumn{2}{c|}{Experiment}  & \multicolumn{2}{c|}{noDA}  &  \multicolumn{2}{c|}{MMD} &  \multicolumn{2}{c|}{MMD + TL} \\\cline{2-9}
  & \multicolumn{2}{c|}{True Label}  & M  &  NM & M &  NM &  M  &  NM   \\\Cline{1.2pt}{2-9}
\multirow{2}{*}{Source} & \multirow{4}{*}{Predicted Label} &  M  &  $0.92$   & $0.08$   &  $0.95$   & $0.07$     & $0.74$   & $0.09$       \\
  &  &  NM  &  $0.08$  & $0.92$    &  $0.05$  & $0.93$    & $0.26$   &   $0.91$  \\\cline{3-9}
\multirow{2}{*}{Target} &  &  M  &  $0.80$    & $0.81$    & $0.63$   & $0.58$   & $0.74$  &  $0.37$ \\
  & &  NM     &  $0.20$    & $0.19$    & $0.37$   &  $0.42$  & $0.26$   & $0.63$      \\\cline{2-9}

\end{tabular}
\end{minipage}
\end{table*}

\begin{table*}
   \centering
   \noindent\begin{minipage}[b]{0.99\textwidth}
   \centering
    \caption{Results from running DeepMerge simulated-to-real experiments with ten different random seeds. Seeds are used for image shuffling, weight initialization, and CUDA backend. We present means and standard deviations for all aforementioned performance metrics.}
  \label{table:seeds_rs}
  \centering
  \begin{tabular}{|l | l | c c| }
    \multicolumn{1}{c}{}  & \multicolumn{3}{c}{DeepMerge: Simulated-to-Real}\\\hline 
Experiment    &  Metric     &  Source   &   Target  \\\Cline{1.2pt}{1-4}
\multirow{5}{*}{No Domain Adaptation}           &  AUC          &   $0.97\pm0.006$       &   $0.52\pm0.04$  \\
                                                &  Accuracy     &    $0.92\pm0.01$       &   $0.50\pm0.02$  \\ 
                                                &  Precision    &    $0.94\pm0.02$        &  $0.50\pm0.02$    \\
                                                &  Recall       &    $0.90\pm0.03$        &  $0.94\pm0.06$ \\
                                                &  F1 score     &    $0.92\pm0.01$        &  $0.65\pm0.02$     \\
                                                &  Brier score  &    $0.06\pm0.01$        &  $0.50\pm0.01$  \\\hline
\multirow{5}{*}{MMD}                            &  AUC          &     $0.98\pm0.006$      &  $0.61\pm0.04$     \\
                                                &  Accuracy     &     $0.94\pm0.01$      &  $0.55\pm0.04$   \\ 
                                                &  Precision    &     $0.95\pm0.02$      &  $0.55\pm0.04$    \\
                                                &  Recall       &     $0.93\pm0.04$      &   $0.63\pm0.06$    \\
                                                &  F1 score     &    $0.94\pm0.02$       &  $0.58\pm0.02$    \\
                                                &  Brier score  &   $0.05\pm0.01$       &   $0.38\pm0.04$     \\\hline
\multirow{5}{*}{MMD + Transfer Learning}         &  AUC          &   $0.84\pm0.05$        &  $0.67\pm0.05$     \\
                                                &  Accuracy     &      $0.78\pm0.05$     &     $0.62\pm0.04$    \\ 
                                                &  Precision    &     $0.85\pm0.07$      &   $0.62\pm0.05$    \\
                                                &  Recall       &      $0.67\pm0.05$     &   $0.64\pm0.06$      \\
                                                &  F1 score     &    $0.75\pm0.06$       &  $0.63\pm0.04$     \\
                                                &  Brier score  &       $0.16\pm0.03$    &  $0.27\pm0.02$      \\\hline
\end{tabular}
\end{minipage}
\end{table*}


\bsp	
\label{lastpage}
\end{document}